\documentclass[aps,prl,reprint,twocolumn,superscriptaddress,longbibliography]{revtex4-2}
\usepackage{amsmath}
\usepackage{amsfonts}
\usepackage{MnSymbol}
\usepackage{hyperref}
\hypersetup{
	colorlinks=true,
	linkcolor=blue,
	filecolor=blue,
	citecolor = blue,      
	urlcolor=cyan,
}

\usepackage{xcolor}
\usepackage{graphicx}
\usepackage{siunitx}
\newcommand{\figref}[1]{Fig.~\ref{#1}}

\newcommand{\eqeqref}[1]{Eq.~(\ref{#1})}

\newcommand{\Ket}[1]{\left|#1\right>}

\newcommand{\ExpVal}[1]{\langle #1\rangle}

\begin{document}

\title{Entangled states from sparsely coupled spins for metrology with neutral atoms}

\author{Sridevi Kuriyattil}
\email{kuriyattil-sridevi@strath.ac.uk}
\author{Pablo M. Poggi}
\author{Jonathan D. Pritchard}
\author{Johannes Kombe}
\affiliation{Department of Physics, SUPA and University of Strathclyde, Glasgow G4 0NG, United Kingdom}
\author{Andrew J. Daley}
\affiliation{Department of Physics, SUPA and University of Strathclyde, Glasgow G4 0NG, United Kingdom}
\affiliation{Clarendon Laboratory, University of Oxford, Parks Road, Oxford OX1 3PU, United Kingdom}

\begin{abstract}
Quantum states featuring extensive multipartite entanglement are a resource for quantum-enhanced metrology, with sensitivity up to the Heisenberg limit. However, robust generation of these states using unitary dynamics typically requires all-to-all interactions among particles. Here, we demonstrate that optimal states for quantum sensing can be generated with sparse interaction graphs featuring only a logarithmic number of couplings per particle.  
We show that specific sparse graphs with long-range interactions can approximate the dynamics of all-to-all spin models, such as the one-axis twisting model, even for large system sizes. 
The resulting sparse coupling graphs and protocol can also be efficiently implemented using dynamic reconfiguration of atoms in optical tweezers.
\end{abstract}
\maketitle
Entangled resource states are the foundation of quantum metrology, enabling quantum systems to surpass classical limits of measurement precision. This has been extensively studied, beginning with realisations of squeezed states, which typically rely on symmetric all-to-all interactions \cite{Kitagawa_1993,Wineland_1994,ma_quantum_2011,Wineland_1992,Braverman_PRL2019,esteve2008,gross2010,Li_PRX2022,Colombo2022,hines2023}.  Recently, it has been demonstrated that neutral atom arrays can be used to implement a wide range of long-range coupling graphs \cite{Bluvstein_2023,Bluvstein_2022, Xu2024,Bornet_2023,eckner2023}. Beyond their role as a versatile platform for quantum computing \cite{Brennen_1999_QL,Jaksch_2000_Fast_quantum,Brion_2007_QCE,Mølmer_2011,Saffman_2016,Weiss_2017, Adams_2020, Henriet2020quantumcomputing,levine2019parallel,CongHardware2022,wu2022erasure,AugerBlueprint2017,Jaksch_2000_Fast_quantum,Saffman_2016,Bluvstein_2023, Bluvstein_2022,Henriet2020quantumcomputing, Evered2023, Xu2024,nikolov2023,Pelegri_2022,Jandura_2022,Pagano_2022,fromonteil2024} and simulation \cite{Bernien_2017,Browaeys_2020,Omran_2019,Veressen_2021,Semeghini_2021,Labuhn2016}, these systems hold significant promise for next-generation atomic clocks and broader applications in metrology \cite{CampbellYe2017, HutsonYe2019, BACON_collab_2021,eckner2023,hines2023,bornet2023}. In this work, we show that sparse non-local coupling graphs produced by shuffling operations in neutral atom arrays can be used to prepare highly entangled states as a resource for quantum metrology, by emulating one axis twisting with all-to-all connectivity with only  $\sim \log_{2}(N)$ couplings per particle, where $N$ is the system size.

\begin{figure}[] 
        \centering     {\includegraphics[width=1.00\linewidth]{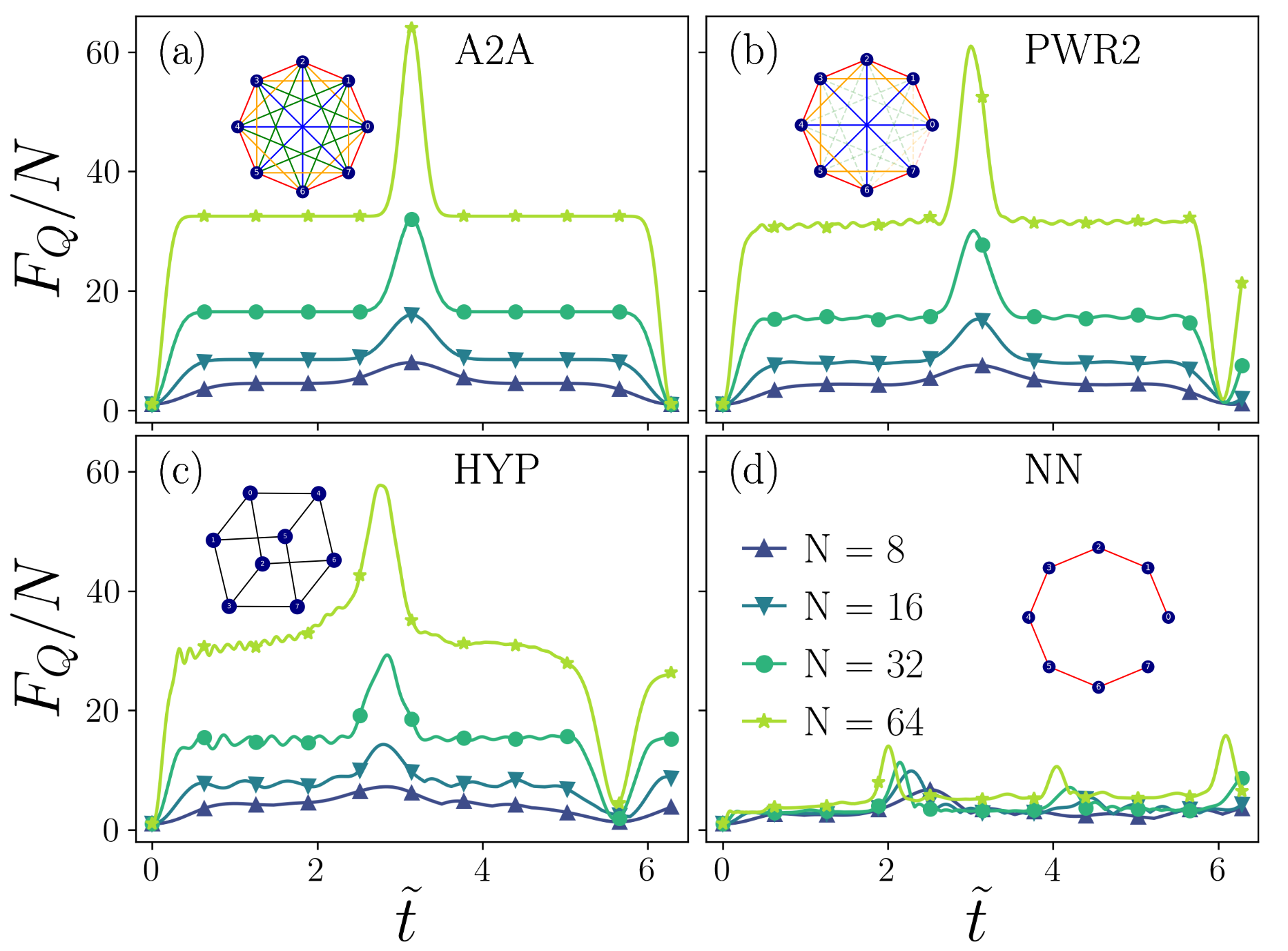}}
        \caption{
        (a)-(d) Evolution of the Quantum Fisher Information (QFI) density $F_Q/N$ under the $XY$ Hamiltonian given in \eqeqref{Eq:XY_Hamiltonian} for different coupling graphs $\chi_{ij}$ and system sizes $N=8, 16, 32, 64$ (see legend in (d) for colours and markers). The spins are initialised in an $x$-polarized state and the time is normalised according to \eqeqref{eq:time_norm}. The dynamics of the sparse coupling graphs PWR2 and hypercube closely resemble those of the A2A coupling graph, leading to a QFI density that scales with system size. On the other hand, the NN coupling graph leads to a significantly lower QFI. Insets in each panel display illustrations of the respective coupling graphs. The computations are performed with 1000 data points. For clarity of the figure only a subset of the computed data points are shown.
        }
        \label{Fig1:Main_Result}
\end{figure}

The models we consider are inspired by  sparse coupling graphs that have been explored in relation to quantum information scrambling both theoretically \cite{Tomohiro_deter_2021, Greg_2019_tree,Gubser_2018_continuum,Gubser_2019_mixed,Kuriyattil_2023,Hashizume_2022} and experimentally \cite{periwal_programmable_2021,Barredo_2016, Kim_2016, Manuel_atom_assembly_2016, Bluvstein_2022}. We compare and contrast two sparse coupling models with an all-to-all (A2A), and a nearest neighbour (NN) coupling geometry.
In the \textit{Powers of Two} (PWR2) model, spins interact if and only if separated by a distance equal to an integer power of 2. This model's coupling graph has $\mathcal{E}_{G} = N \log_{2}(N) -N + 1$ edges for $N$ vertices. In contrast, for the \textit{Hypercube coupling graph} spins reside on the vertices of a $m=\log_{2}(N)$ dimensional hypercube, resulting in $\mathcal{E}_{G} = \frac{N}{2}\log_{2}(N)$ edges for $N$ vertices. Illustrations of the respective coupling graphs for $N=8$ are presented in \figref{Fig1:Main_Result}.

\begin{figure}[] 
        \centering{\includegraphics[width=1.00\linewidth]{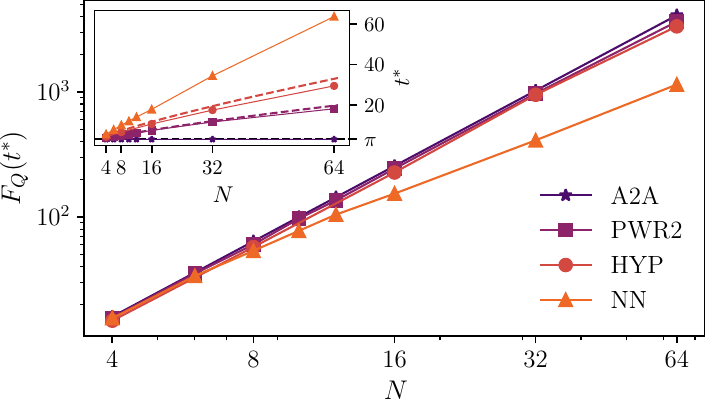}}
        \caption{ The maximum value of the QFI under the dynamics generated by the $XY$ Hamiltonian in \eqeqref{Eq:XY_Hamiltonian} on an initially $x$ polarized state is shown as a function of system size $N$ for different coupling graphs. We extract the scaling coefficients $F_Q \propto N^{\beta}$ for different coupling graphs, yielding $\beta = 2.00$ for A2A coupling, $\beta = 1.96$ for the hypercube, and $\beta = 1.97$ for the PWR2 graph. The NN coupling shows a much weaker scaling of $ \beta = 1.51$. (inset) Physical time $t^{*}$ to reach the maximum QFI as a function of $N$.
        Markers show numerically obtained values, while dashed lines indicate the expected scaling based on the mean-field model $t^{*} \approx \pi \left(\frac{\mathcal{E}_{G,A2A}}{\mathcal{E}_{G}} \right)$.}
        
        \label{Fig2:Scaling}
\end{figure}

Below, we explore how well the dynamics generated by these types of interactions approximate the one-axis twisting (OAT) Hamiltonian, a well-studied model which features uniform, infinite-range Ising interactions \cite{Kitagawa_1993,ma_quantum_2011,munoz2023}. In the OAT Hamiltonian, all $N$ spin-$1/2$ particles interact with each other, resulting in a dense, \textit{all-to-all} connectivity with $\mathcal{E}_{G,A2A} = N(N-1)/2$ edges as shown in \figref{Fig1:Main_Result}(a). While these can be engineered indirectly (e.g, via photon-mediated interactions between atoms in an optical cavity \cite{Britton_2012,Justin_qsd_2016,Emily_Photon_mdeiated_2019,Ritsch_cold_atoms_2013,Baumann_2010}), this degree of connectivity is challenging to realize with direct interactions in most physical setups. In the following, we show that the all-to-all dynamics can be well approximated by sparse coupling graphs, and propose to use shuffling operations, interleaved with nearest-neighbour interactions and local rotations, to emulate the OAT dynamics efficiently.
\\
\paragraph{Dynamical preparation of entangled states using sparse coupling graphs.} We consider spins initially prepared in an $x$-polarized state $|+ \rangle^{\otimes N}$ and allow it to evolve under the $XY$ Hamiltonian,
\begin{align}
\label{Eq:XY_Hamiltonian}
H_{XY} = \sum_{i,j}\chi_{ij}(S_{i}^{x}S_{j}^{x} +S_{i}^{y}S_{j}^{y}) ~,
\end{align}
for different coupling graphs $\{\chi_{ij}\}$, where the spin operators in terms of Pauli operators are given by $S_{i}^{\alpha} = \sigma_{i}^{\alpha}/2$ ($\hbar = 1$), and $\alpha\in\{x,y,z\}$. The choice of the Hamiltonian is inspired by the studies conducted in \cite{perlin2020,block2023universal}. 
To account for different numbers of bonds in the different coupling graphs considered here, we introduce the normalized time
\begin{align}
    \label{eq:time_norm}
    \tilde{t} = t\frac{\mathcal{E}_{G}}{\mathcal{E}_{G,A2A}} ~,
\end{align}
where $\mathcal{E}_{G}$ is the number of edges in a given graph $G$, and $t$ is the physical time. For the A2A graph, $\chi_{ij}=\chi_0$ for $i\neq j$, and \eqeqref{Eq:XY_Hamiltonian} is proportional to the OAT Hamiltonian $H_{\text{OAT}}=\chi_0 J_z^2$, where $J_z=\sum_i S_i^{z}$.
\\
We quantify the generation of metrologicaly-useful entanglement via the Quantum Fisher Information (QFI) \cite{PezzeTreutlein2018, PezzeSmerzi2014, BraunsteinCaves1994, Paris2009, TothApellaniz2014, Helstrom1969} of a quantum state $\Ket{\psi}$ which determines the optimal precision with which a system parameter $\theta$ can be estimated using $\Ket{\psi}$ as a probe. 
For spin systems, the standard scenario corresponds to $\theta$ being a phase encoded by a small rotation of a pure probe state around a given axis. 
The QFI in such a scenario is
\begin{align}
    \label{Eq:QFI}
    F_Q[\theta, J_\alpha] = 4 (\Delta J_\alpha)^2 ~,
\end{align}
where $J_\alpha=\sum_i S_i^\alpha$ is the generator of a rotation around axis $\alpha$ with $\alpha=x,y,z$, and  and $(\Delta J_\alpha)^2=\ExpVal{J_\alpha^2} - \ExpVal{J_\alpha}^2$ is its variance calculated over the state $\Ket{\psi}$.
For uncorrelated particles, the sensitivity to rotations is bounded by classical limits on measurement precision which yield $F_Q = N$ (standard quantum limit), where $N$ is the number of particles in the system. However, the presence of entanglement in the system pushes the sensitivity beyond this, indicating utility in phase estimation and metrology, with $F_{Q} = N^{2}$ being the fundamental Heisenberg limit in the absence of noise \cite{PezzeSmerzi2014}.
\\
In \figref{Fig1:Main_Result}, we show the evolution of the QFI for different coupling graphs and system sizes. The dynamics of the QFI for the PWR2 and hypercube coupling graphs, \figref{Fig1:Main_Result}(b) and (c), are reminiscent of the dynamics of the A2A coupling graph, \figref{Fig1:Main_Result}(a). In all three cases, at $t \lesssim 1 /\sqrt{N}$ there is an initial spin-squeezed region where the QFI rises steadily, followed by a plateau region, and a final rise to reach a state with maximum QFI, $F_{Q} \simeq N^{2}$, at $\tilde{t} = \pi$. 
To support this statement, we have calculated the squeezing parameter, which quantifies metrologically useful entanglement, at $\tilde{t}=1/\sqrt{N}$, the QFI at the plateau region, and the overlap of the state with maximum QFI with the GHZ state, as detailed in the Supplementary Material \cite{SupMat}. To understand the gain we obtain from using sparse interactions, it is instructive to compare the dynamics to the nearest neighbour model, \figref{Fig1:Main_Result}(d), which shows much lower QFI throughout the dynamical evolution. Importantly, the QFI per particle fails to scale with system size, thus signifying no scaling gain with respect to the standard quantum limit. It is thus evident that the sparse models create states with significantly higher QFI and therein emulate the all-to-all OAT dynamics.
\\
In \figref{Fig2:Scaling} we analyse how the maximum QFI scales with system size. Specifically, we propose an ansatz $F_Q(t^{*}) \propto N^{\beta}$, where $t^{*}$ is the physical time at which the QFI reaches its maximum, and extract $\beta$ numerically. We find that the PWR2 and hypercube geometries lead to states that follow $F_Q \propto N^{1.97}$ and $F_Q \propto N^{1.96}$ respectively. While we employ exact diagonalisation for small system sizes, all results for $N>16$ are obtained using the time-dependent variational principle (TDVP) approach with matrix product states (MPS) \cite{WhiteFeiguin2004, DaleySchollwoeck2004, Schollwoeck2011, HaegemanVerstraete2016, FishmanStoudenmire2022, ITensor2022}. The main text shows the results for bond dimension $D=256$ for $N=32$ and $D=512$ for $N=64$, truncation error $\epsilon = 10^{-13}$, and the time step $\chi_{0}dt=10^{-2}$. We refer the reader to the Supplementary Material \cite{SupMat} for a detailed analysis of the numerical convergence.
The inset of \figref{Fig2:Scaling} illustrates the dependence of $t^{*}$ on the system size $N$. For the all-to-all coupling graph, this time is known to be constant at $t^{*}=\pi$ \cite{PezzeTreutlein2018}. For the NN model, as expected, it increases linearly with the system size, whereas we observe that $t^{*} \sim N/\log_{2}(N)$ in the sparse models. This scaling is consistent with the mean-field prediction given by Eq. (\ref{eq:time_norm}), which is shown as dashed lines in the figure.



\begin{figure}[t!] 
        \centering
        \hfill        {\includegraphics[width=1.00\linewidth]{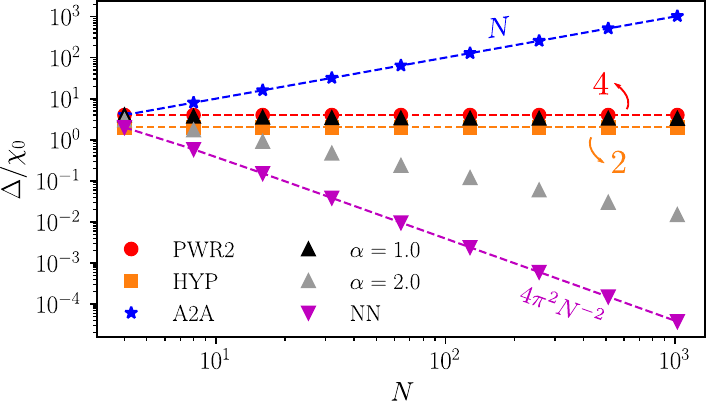}}
        \caption
        {Spectral gap $\Delta/\chi_0$ of the isotropic Heisenberg Hamiltonian in Eq. (\ref{eq:gap_goat}) as a function of system size $N$ for different coupling graphs with various levels of connectivity. Systems with asymptotically vanishing gaps deviate from the collective OAT dynamics, while systems with non-vanishing gaps in the large-$N$ limit display robust collective dynamics that can generate metrologically useful states. Examples of the former are the nearest-neighbour graph (NN) and the power-law coupling graph with $\alpha>1$. Meanwhile, the latter includes the all-to-all (A2A) and power-law couplings with $\alpha \leq 1$ (for $1$D graphs), as well as the sparse coupling graphs PWR2 and Hypercube studied here, which display a constant gap as a function of system size. In all cases, symbols indicate numerical evaluation of the gap from an exact expression (see Supplemental material \cite{SupMat}), while dashed lines indicate closed-form analytical approximations, indicated for each case in the figure.}
        \label{Fig3:Gap}
\end{figure}

\paragraph{Sparse coupling graphs and collective behavior} We can analyse the results obtained from the sparse coupling graphs by studying how robust the OAT dynamics are to perturbations that break the permutational invariance of the Hamiltonian. 
To this end, it is instructive to rewrite the $XY$  Hamiltonian in \eqeqref{Eq:XY_Hamiltonian} for a given coupling graph $\{\chi_{ij}\}$ as the sum of
\begin{flalign}
  && H_{{\rm gOAT}} &= \chi_0 J_z^2 - \sum\limits_{ij} \chi_{ij} \vec{S}_i.\vec{S}_j ~, \label{eq:gap_goat} &\\
   \text{and} && V_{\rm pert} &= - \sum\limits_{ij} (\chi_0-\chi_{ij})S_i^z S_j^z ~. &
\end{flalign}

Here, $H_{{\rm gOAT}}$ is a generalized OAT model (introduced in Ref. \cite{perlin2020}) composed of the usual OAT twisting term and the isotropic Heisenberg Hamiltonian. Both of these commute with the collective spin operator $\mathbf{J}^2= J_x^2+J_y^2+J_z^2$ for any coupling graph, thus restricting the dynamics of initial spin coherent states to the subspace of permutationally symmetric states. The last term acts as a perturbation that will couple states of different total $\mathbf{J}$, causing the state to leave the symmetric subspace and deviate from the OAT dynamics. Perturbation theory indicates that such deviation will be more prominent if the eigenstates of $H_{{\rm gOAT}}$ corresponding to different values of $\mathbf{J}$ are separated by a small gap. Thus, here we study the scaling of the spectral gap $\Delta$ in the gOAT Hamiltonian with system size for different coupling graphs. In the Supplemental Material \cite{SupMat}, we construct the excitation spectra for the two sparse coupling graphs considered here (PWR2 and Hypercube) using spin-wave theory \cite{Pires_spinwavetheory,stancel2009spin,kittel1963quantum}.
For comparison, we also include results for the well-known case of algebraically-decaying (AD) interactions where $\chi_{ij}=\chi_0 |i-j|^{-\alpha}$ in 1D (with $\alpha=0$ corresponding to the A2A case, and $\alpha\rightarrow \infty$ to the NN case). In \figref{Fig3:Gap}, we show the behavior of the spectral gap $\Delta$ as a function of system size $N$ for various cases of interest. We observe a power-law scaling $\Delta \sim N^\gamma$ for all coupling graphs, consistent with previous studies. Crucially, we show analytically in the Supplemental Material \cite{SupMat} that the sparse graphs lead to constant gaps, i.e. $\Delta/\chi_0=2,4$ for the hypercube and PWR2 respectively. This implies that the collective dynamics generated by these models are well protected from the non-collective perturbations, thus explaining why these models generate metrologically useful states. 
Furthermore, since the gap remains exactly constant (instead of increasing with $N$), we argue that this level of connectivity is optimal for producing such behavior. It closely mimics the behavior of the known optimal case of AD couplings with $\alpha=1$ for 1D graphs.

\begin{figure}[t] 
        \centering
        \hfill
        {\includegraphics[width=1.00\linewidth]{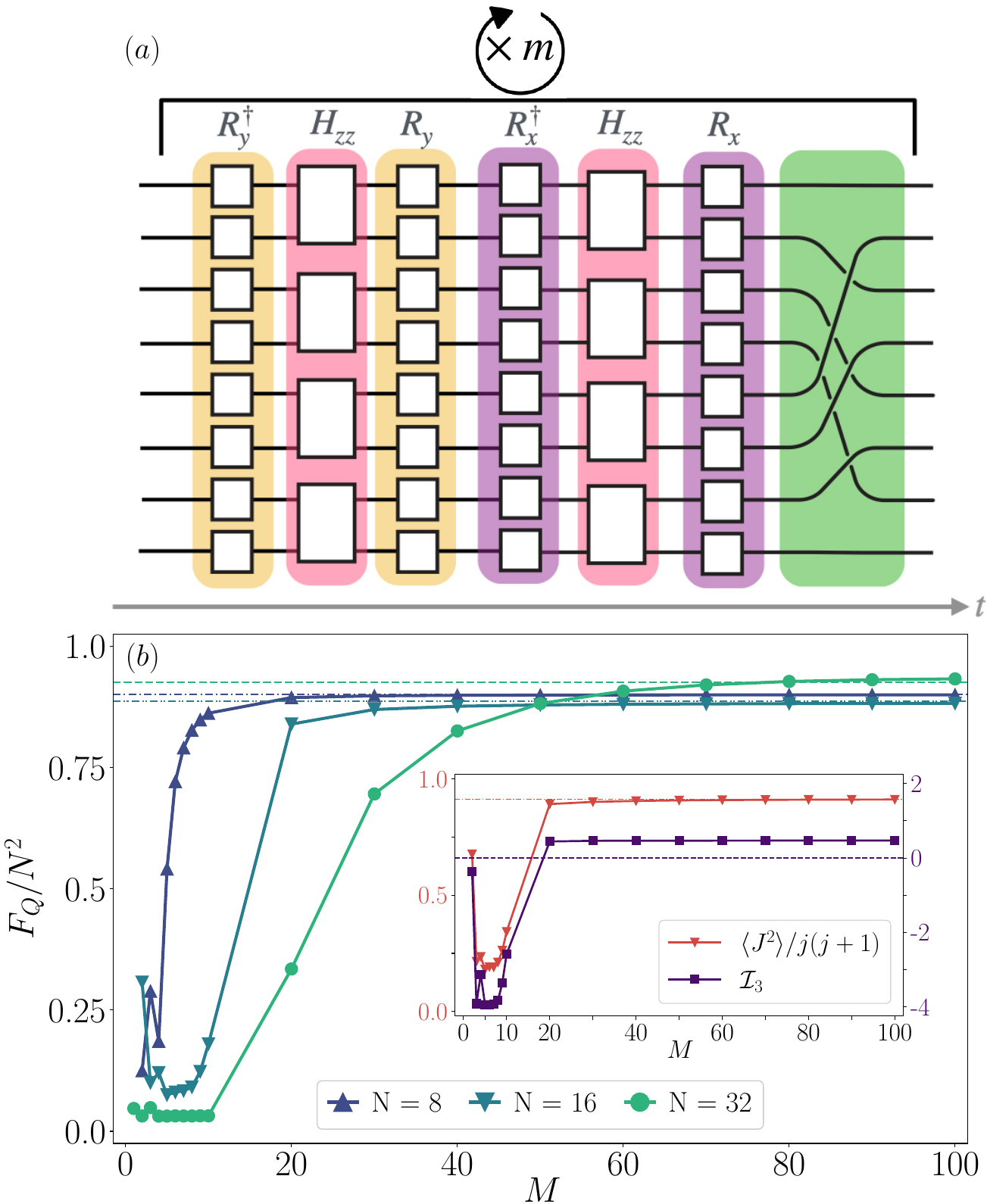}}
        \caption{
        (a) Our circuit implementation consists of single qubit rotations \eqeqref{Eq:single_qubit_operations} (purple for $R_x$ and yellow for $R_y$) followed by $zz$ interactions between nearest neighbour spins according to Hamiltonian in \eqeqref{Eq:ZZ_interaction} (pink for $H_{zz}$) and inverse Faro shuffles $\mathcal{R}$ (green). The application of $H_{zz}$ and $R_\alpha$ are done in $m=\log_{2}(N)$ steps, and leads to a first-order Trotter decomposition of the spin-interaction Hamiltonian according to \eqeqref{Eq:Strob_evolve}.
        (b) The normalised QFI $F_{Q} / N^{2}$ as a function of the number of iterations $M$, where we evolve the initial $x$-polarized state up to $t^{*}$ for the hypercube geometry. The dashed lines of corresponding colours represent the values of $F_{Q}/N^{2}$ at $t=t^{*}$ for $N=\{8,16,32\}$ respectively, extracted from the continuous time evolution. (inset) The expectation value of the collective spin operator $\langle J^2 \rangle$ normalised to $j(j+1)$, with $j=N/2$  (red, left axis), and $\mathcal{I}_{3}$ (violet, right axis) are shown as a function of $M$ for $N=16$. For small values of $M$, there is a strong deviation from the collective spin behavior, and a loss of permutational symmetry indicating a scrambled state characterized by negative $\mathcal{I}_3$. The horizontal dashdotted (red), and dashed (violet) lines represent $\max\{\langle J^2 \rangle\}$, and  $\mathcal{I}_{3} = 0$ respectively.}
        \label{Fig4:Experimental}
\end{figure}

\paragraph{Implementing sparse graphs in tweezer arrays.} With an analytical understanding of why sparse models preserve collective spin behavior and are able to generate metrologically-relevant resource states, we now shift focus to proposing experimental implementations. Specifically, we focus on neutral atom arrays employing tweezer-assisted shuffling operations. This choice is due to its good scalability, coherent control of each atom, and flexibility in atom movement. 
To realize the hypercube connectivity, these rearrangement operations can be chosen to execute a  ``Faro-shuffle'' \cite{aldous1986shuffling,diaconis1983mathematics}, which moves the particle originally located in site $i$ to site $i'$ according to the following rule 
\begin{align}
    i'=\mathcal{R}(i=b_{m}...b_{2}b_{1})=b_{1}b_{m}...b_{2} ~.
    \label{Eq:Faro_Shuffle}
\end{align}
Hereby, the binary representation of site $i=b_{m}...b_{2}b_{1}$ is reversed, such that the least significant bit $b_{1}$ of $i$ becomes the most significant bit of $i'=\mathcal{R}(i)$. In \cite{SupMat} we propose an alternative shuffling sequence which yields the PWR2 geometry.
To realize the spin-exchange Hamiltonian, we leverage the hypercube geometry native to this Faro-shuffle and a sequence of global rotations and $zz$ (Ising) interactions, 
\begin{flalign}
    \label{Eq:single_qubit_operations}
    && R_{\alpha} &= e^{-i\pi/4\sum_{i}\sigma_{i}^{\alpha}} ~ \alpha \in \{x,y\} ~, & \\
 && 
    \label{Eq:ZZ_interaction}
    H_{zz} &= 2\chi_{0} \sum_{\nu} S_{2\nu-1}^{z}S_{2\nu}^{z} ~, &
\end{flalign}
where $\nu$ indicates the atomic position, rather than the spin index. 
While spin-exchange Hamiltonians can be implemented natively in Rydberg arrays \cite{Bornet_2023}, this requires encoding in highly excited states, which are prone to decay. Our approach, based on hyperfine qubits and shuffling operations, enables robust entanglement generation using spins encoded in long-lived atomic ground states except during short pulsed gate operations.
In \cite{SupMat} we provide details about how to implement this procedure in practice. 
During each time step of duration $dt = t^{*}/M$, we apply $m \in \{ 1,\ldots,\log_{2}(N)\}$ Faro shuffles to build up the hypercube coupling graph. Before each shuffling operation, we evolve the system according to
\begin{align}
    \label{Eq:Strob_ZZ_Hamiltonian}
    U_{m}(dt) = \Big[R_{x} e^{-iH_{zz}dt/2}R_{x}^{\dagger}\Big] \Big[R_{y} e^{-iH_{zz}dt/2}R_{y}^{\dagger}\Big] ~,
\end{align}
which effectively realises a first-order Trotter decomposition, such that the full evolution during one time step takes the following form
\begin{equation}
\label{Eq:Strob_evolve}
    U(t, t+dt) = \prod_{m} \mathcal{R} U_{m}(dt) ~,
\end{equation}
which is schematically depicted in Fig. \ref{Fig4:Experimental} (a).
In \figref{Fig4:Experimental} (b) we show the value of the maximum QFI $F_Q$ achieved by the stroboscopic protocol as a function of the number of iterations $M$, for different system sizes. For large enough $M$, the protocol reaches Heisenberg scaling $F_Q\sim N^2$, as expected from the fine-grained Trotterization. For small $M$, we observe a transition towards smaller values of QFI closer to the standard quantum limit. In this regime, the unitary in \eqeqref{Eq:Strob_ZZ_Hamiltonian} fails to approximate the continuous evolution, and previous studies \cite{Heyl_2019,chinni2022} suggest the existence of a chaotic regime, leading to highly scrambled states with low QFI. We further analyse this behavior by calculating the expectation value of the collective spin operator $\langle \mathbf{J}^2 \rangle$, shown in the inset of \figref{Fig4:Experimental} (b) for $N=16$. For collective spin states $\langle \mathbf{J}^2 \rangle = j(j+1)$, where $j=N/2$, and we see how this collective behavior is lost for $M<10$ iterations.
\\
In addition to these quantities, we also calculate the tripartite mutual information (TMI) $\mathcal{I}_{3}$, which has been extensively studied in the context of information scrambling. The TMI $\mathcal{I}_{3}$ between three regions $A$, $B$, $C$ is defined as 
\begin{align}
    \label{Eq:TMI}
    I(A:B:C) = \mathcal{I}_{3} = I(A;B) + I(A;C) - I(A;BC) ~,
\end{align}
where $I(A;B)= S_{A} + S_{B} - S_{AB}$ is the mutual information between subregions $A$ and $B$, and $S_{A}$ is the von Neumann entropy. A negative $\mathcal{I}_{3}$ signifies that the combined region $BC$ contains more information about subregion $A$ than the subregions $B$ and $C$ individually. Thus negativity of the tripartite mutual information indicates many-body entanglement in the system \cite{hosurChaosQuantumChannels2016,gullans2020dynamical,zabalo2020critical,Kuriyattil_2023}. However, for permutationally symmetric states (such as the collective states considered here) the tripartite mutual information is positive \cite{Seshadri_2018}. A simple example is the Greenberger-Horne-Zeilinger GHZ state, that has $\mathcal{I}_{3} = \ln{2}$. For large $M$, we obtain that the tripartite mutual information is positive as shown in the inset of \figref{Fig4:Experimental} (b).
\\
We observe the presence of a sharp threshold as a function of the number of iterations $M$ that separates permutationally symmetric states at larger $M$ from scrambled states at smaller $M$ characterized by negative $\mathcal{I}_3$. The minimum number of iterations required to attain Heisenberg-limited states increases with system size. While the actual scaling is challenging to analyze numerically because usual MPS methods stop being efficient for scrambled states, we observe a sub-linear scaling of $M$ with the size of the system. Moreover, in the Supplementary Material \cite{SupMat} we show that achieving states featuring Heisenberg scaling (i.e. $F_Q\propto N^2$ rather than $F_Q=N^2)$ is less demanding since shorter evolution times suffice.
\paragraph{Summary and Outlook.}
We demonstrated the generation of resource states for quantum-enhanced metrology from sparse coupling graphs by showing how they effectively emulate one-axis twisting (OAT) dynamics with fewer resources. We also proposed a stroboscopic protocol, implementable on current neutral atom arrays, which requires only tweezer shuffling, nearest-neighbour Ising interactions, and local rotations to prepare metrologically-relevant states exhibiting Heisenberg scaling. While sparse nonlocal graphs had previously been shown to lead to fast information scrambling, our findings demonstrate that tasks requiring states with a specific entanglement structure, such as quantum sensing, can also be improved by the use of this tool. With advancements in atom shuffling using tweezer arrays \cite{Bluvstein_2022,Bluvstein_2023,Barredo_2016}, squeezing generated by these sparse graphs could potentially be applied in optical tweezer clocks, paving the way for entanglement-enhanced optical clocks \cite{Matthew_opticalclocks,Madjarov2019_opticalclocks,eckner2023}. Our results pave the way for further exploration of the utility of sparse coupling graphs for other tasks in quantum computing and quantum simulation. In particular, combining sparse nonlocal interactions within the framework of variational circuits has the potential to open new avenues for the generation of useful entangled states for quantum optimization and quantum machine learning \cite{cerezo2021}. Moreover, extending these techniques to digital circuit models could open new avenues for fault-tolerant preparation of metrologically relevant states. 
\begin{acknowledgements}
PMP acknowledges insightful discussions with Stephen Piddock about the spectral analysis of the Heisenberg model. We are grateful to Gregory Bentsen for valuable inputs on the scrambling properties of sparse graphs. We also thank Gerard Pelegrí for valuable insights into experimental implementations. This work was supported by EPSRC through grant number EP/Y005058/2 and by the Programme grant QQQS (Grant No. EP/Y01510X/1), as well as the EPSRC Prosperity Partnership \emph{SQuAre} (Grant No. EP/T005386/1) with funding from M Squared Lasers Ltd. The data presented in this work are available at \cite{data_open_access}. 
\end{acknowledgements}
\bibliography{References.bib}

\begin{thebibliography}{90}%
\makeatletter
\providecommand \@ifxundefined [1]{%
 \@ifx{#1\undefined}
}%
\providecommand \@ifnum [1]{%
 \ifnum #1\expandafter \@firstoftwo
 \else \expandafter \@secondoftwo
 \fi
}%
\providecommand \@ifx [1]{%
 \ifx #1\expandafter \@firstoftwo
 \else \expandafter \@secondoftwo
 \fi
}%
\providecommand \natexlab [1]{#1}%
\providecommand \enquote  [1]{``#1''}%
\providecommand \bibnamefont  [1]{#1}%
\providecommand \bibfnamefont [1]{#1}%
\providecommand \citenamefont [1]{#1}%
\providecommand \href@noop [0]{\@secondoftwo}%
\providecommand \href [0]{\begingroup \@sanitize@url \@href}%
\providecommand \@href[1]{\@@startlink{#1}\@@href}%
\providecommand \@@href[1]{\endgroup#1\@@endlink}%
\providecommand \@sanitize@url [0]{\catcode `\\12\catcode `\$12\catcode `\&12\catcode `\#12\catcode `\^12\catcode `\_12\catcode `\%12\relax}%
\providecommand \@@startlink[1]{}%
\providecommand \@@endlink[0]{}%
\providecommand \url  [0]{\begingroup\@sanitize@url \@url }%
\providecommand \@url [1]{\endgroup\@href {#1}{\urlprefix }}%
\providecommand \urlprefix  [0]{URL }%
\providecommand \Eprint [0]{\href }%
\providecommand \doibase [0]{https://doi.org/}%
\providecommand \selectlanguage [0]{\@gobble}%
\providecommand \bibinfo  [0]{\@secondoftwo}%
\providecommand \bibfield  [0]{\@secondoftwo}%
\providecommand \translation [1]{[#1]}%
\providecommand \BibitemOpen [0]{}%
\providecommand \bibitemStop [0]{}%
\providecommand \bibitemNoStop [0]{.\EOS\space}%
\providecommand \EOS [0]{\spacefactor3000\relax}%
\providecommand \BibitemShut  [1]{\csname bibitem#1\endcsname}%
\let\auto@bib@innerbib\@empty
\bibitem [{\citenamefont {Kitagawa}\ and\ \citenamefont {Ueda}(1993)}]{Kitagawa_1993}%
  \BibitemOpen
  \bibfield  {author} {\bibinfo {author} {\bibfnamefont {M.}~\bibnamefont {Kitagawa}}\ and\ \bibinfo {author} {\bibfnamefont {M.}~\bibnamefont {Ueda}},\ }\href {https://doi.org/10.1103/PhysRevA.47.5138} {\bibfield  {journal} {\bibinfo  {journal} {Phys. Rev. A}\ }\textbf {\bibinfo {volume} {47}},\ \bibinfo {pages} {5138} (\bibinfo {year} {1993})}\BibitemShut {NoStop}%
\bibitem [{\citenamefont {Wineland}\ \emph {et~al.}(1994)\citenamefont {Wineland}, \citenamefont {Bollinger}, \citenamefont {Itano},\ and\ \citenamefont {Heinzen}}]{Wineland_1994}%
  \BibitemOpen
  \bibfield  {author} {\bibinfo {author} {\bibfnamefont {D.~J.}\ \bibnamefont {Wineland}}, \bibinfo {author} {\bibfnamefont {J.~J.}\ \bibnamefont {Bollinger}}, \bibinfo {author} {\bibfnamefont {W.~M.}\ \bibnamefont {Itano}},\ and\ \bibinfo {author} {\bibfnamefont {D.~J.}\ \bibnamefont {Heinzen}},\ }\href {https://doi.org/10.1103/PhysRevA.50.67} {\bibfield  {journal} {\bibinfo  {journal} {Phys. Rev. A}\ }\textbf {\bibinfo {volume} {50}},\ \bibinfo {pages} {67} (\bibinfo {year} {1994})}\BibitemShut {NoStop}%
\bibitem [{\citenamefont {Ma}\ \emph {et~al.}(2011)\citenamefont {Ma}, \citenamefont {Wang}, \citenamefont {Sun},\ and\ \citenamefont {Nori}}]{ma_quantum_2011}%
  \BibitemOpen
  \bibfield  {author} {\bibinfo {author} {\bibfnamefont {J.}~\bibnamefont {Ma}}, \bibinfo {author} {\bibfnamefont {X.}~\bibnamefont {Wang}}, \bibinfo {author} {\bibfnamefont {C.~P.}\ \bibnamefont {Sun}},\ and\ \bibinfo {author} {\bibfnamefont {F.}~\bibnamefont {Nori}},\ }\href {https://doi.org/10.1016/j.physrep.2011.08.003} {\bibfield  {journal} {\bibinfo  {journal} {Physics Reports}\ }\textbf {\bibinfo {volume} {509}},\ \bibinfo {pages} {89} (\bibinfo {year} {2011})}\BibitemShut {NoStop}%
\bibitem [{\citenamefont {Wineland}\ \emph {et~al.}(1992)\citenamefont {Wineland}, \citenamefont {Bollinger}, \citenamefont {Itano}, \citenamefont {Moore},\ and\ \citenamefont {Heinzen}}]{Wineland_1992}%
  \BibitemOpen
  \bibfield  {author} {\bibinfo {author} {\bibfnamefont {D.~J.}\ \bibnamefont {Wineland}}, \bibinfo {author} {\bibfnamefont {J.~J.}\ \bibnamefont {Bollinger}}, \bibinfo {author} {\bibfnamefont {W.~M.}\ \bibnamefont {Itano}}, \bibinfo {author} {\bibfnamefont {F.~L.}\ \bibnamefont {Moore}},\ and\ \bibinfo {author} {\bibfnamefont {D.~J.}\ \bibnamefont {Heinzen}},\ }\href {https://doi.org/10.1103/PhysRevA.46.R6797} {\bibfield  {journal} {\bibinfo  {journal} {Phys. Rev. A}\ }\textbf {\bibinfo {volume} {46}},\ \bibinfo {pages} {R6797} (\bibinfo {year} {1992})}\BibitemShut {NoStop}%
\bibitem [{\citenamefont {Braverman}\ \emph {et~al.}(2019)\citenamefont {Braverman}, \citenamefont {Kawasaki}, \citenamefont {Pedrozo-Pe\~nafiel}, \citenamefont {Colombo}, \citenamefont {Shu}, \citenamefont {Li}, \citenamefont {Mendez}, \citenamefont {Yamoah}, \citenamefont {Salvi}, \citenamefont {Akamatsu}, \citenamefont {Xiao},\ and\ \citenamefont {Vuleti\ifmmode~\acute{c}\else \'{c}\fi{}}}]{Braverman_PRL2019}%
  \BibitemOpen
  \bibfield  {author} {\bibinfo {author} {\bibfnamefont {B.}~\bibnamefont {Braverman}}, \bibinfo {author} {\bibfnamefont {A.}~\bibnamefont {Kawasaki}}, \bibinfo {author} {\bibfnamefont {E.}~\bibnamefont {Pedrozo-Pe\~nafiel}}, \bibinfo {author} {\bibfnamefont {S.}~\bibnamefont {Colombo}}, \bibinfo {author} {\bibfnamefont {C.}~\bibnamefont {Shu}}, \bibinfo {author} {\bibfnamefont {Z.}~\bibnamefont {Li}}, \bibinfo {author} {\bibfnamefont {E.}~\bibnamefont {Mendez}}, \bibinfo {author} {\bibfnamefont {M.}~\bibnamefont {Yamoah}}, \bibinfo {author} {\bibfnamefont {L.}~\bibnamefont {Salvi}}, \bibinfo {author} {\bibfnamefont {D.}~\bibnamefont {Akamatsu}}, \bibinfo {author} {\bibfnamefont {Y.}~\bibnamefont {Xiao}},\ and\ \bibinfo {author} {\bibfnamefont {V.}~\bibnamefont {Vuleti\ifmmode~\acute{c}\else \'{c}\fi{}}},\ }\href {https://doi.org/10.1103/PhysRevLett.122.223203} {\bibfield  {journal} {\bibinfo  {journal} {Phys. Rev. Lett.}\ }\textbf {\bibinfo {volume} {122}},\ \bibinfo {pages} {223203} (\bibinfo {year} {2019})}\BibitemShut {NoStop}%
\bibitem [{\citenamefont {Esteve}\ \emph {et~al.}(2008)\citenamefont {Esteve}, \citenamefont {Gross}, \citenamefont {Weller}, \citenamefont {Giovanazzi},\ and\ \citenamefont {Oberthaler}}]{esteve2008}%
  \BibitemOpen
  \bibfield  {author} {\bibinfo {author} {\bibfnamefont {J.}~\bibnamefont {Esteve}}, \bibinfo {author} {\bibfnamefont {C.}~\bibnamefont {Gross}}, \bibinfo {author} {\bibfnamefont {A.}~\bibnamefont {Weller}}, \bibinfo {author} {\bibfnamefont {S.}~\bibnamefont {Giovanazzi}},\ and\ \bibinfo {author} {\bibfnamefont {M.~K.}\ \bibnamefont {Oberthaler}},\ }\href@noop {} {\bibfield  {journal} {\bibinfo  {journal} {Nature}\ }\textbf {\bibinfo {volume} {455}},\ \bibinfo {pages} {1216} (\bibinfo {year} {2008})}\BibitemShut {NoStop}%
\bibitem [{\citenamefont {Gross}\ \emph {et~al.}(2010)\citenamefont {Gross}, \citenamefont {Zibold}, \citenamefont {Nicklas}, \citenamefont {Esteve},\ and\ \citenamefont {Oberthaler}}]{gross2010}%
  \BibitemOpen
  \bibfield  {author} {\bibinfo {author} {\bibfnamefont {C.}~\bibnamefont {Gross}}, \bibinfo {author} {\bibfnamefont {T.}~\bibnamefont {Zibold}}, \bibinfo {author} {\bibfnamefont {E.}~\bibnamefont {Nicklas}}, \bibinfo {author} {\bibfnamefont {J.}~\bibnamefont {Esteve}},\ and\ \bibinfo {author} {\bibfnamefont {M.~K.}\ \bibnamefont {Oberthaler}},\ }\href@noop {} {\bibfield  {journal} {\bibinfo  {journal} {Nature}\ }\textbf {\bibinfo {volume} {464}},\ \bibinfo {pages} {1165} (\bibinfo {year} {2010})}\BibitemShut {NoStop}%
\bibitem [{\citenamefont {Li}\ \emph {et~al.}(2022)\citenamefont {Li}, \citenamefont {Braverman}, \citenamefont {Colombo}, \citenamefont {Shu}, \citenamefont {Kawasaki}, \citenamefont {Adiyatullin}, \citenamefont {Pedrozo-Pe\~nafiel}, \citenamefont {Mendez},\ and\ \citenamefont {Vuleti\ifmmode~\acute{c}\else \'{c}\fi{}}}]{Li_PRX2022}%
  \BibitemOpen
  \bibfield  {author} {\bibinfo {author} {\bibfnamefont {Z.}~\bibnamefont {Li}}, \bibinfo {author} {\bibfnamefont {B.}~\bibnamefont {Braverman}}, \bibinfo {author} {\bibfnamefont {S.}~\bibnamefont {Colombo}}, \bibinfo {author} {\bibfnamefont {C.}~\bibnamefont {Shu}}, \bibinfo {author} {\bibfnamefont {A.}~\bibnamefont {Kawasaki}}, \bibinfo {author} {\bibfnamefont {A.~F.}\ \bibnamefont {Adiyatullin}}, \bibinfo {author} {\bibfnamefont {E.}~\bibnamefont {Pedrozo-Pe\~nafiel}}, \bibinfo {author} {\bibfnamefont {E.}~\bibnamefont {Mendez}},\ and\ \bibinfo {author} {\bibfnamefont {V.}~\bibnamefont {Vuleti\ifmmode~\acute{c}\else \'{c}\fi{}}},\ }\href {https://doi.org/10.1103/PRXQuantum.3.020308} {\bibfield  {journal} {\bibinfo  {journal} {PRX Quantum}\ }\textbf {\bibinfo {volume} {3}},\ \bibinfo {pages} {020308} (\bibinfo {year} {2022})}\BibitemShut {NoStop}%
\bibitem [{\citenamefont {Colombo}\ \emph {et~al.}(2022)\citenamefont {Colombo}, \citenamefont {Pedrozo-Peñafiel}, \citenamefont {Adiyatullin} \emph {et~al.}}]{Colombo2022}%
  \BibitemOpen
  \bibfield  {author} {\bibinfo {author} {\bibfnamefont {S.}~\bibnamefont {Colombo}}, \bibinfo {author} {\bibfnamefont {E.}~\bibnamefont {Pedrozo-Peñafiel}}, \bibinfo {author} {\bibfnamefont {A.~F.}\ \bibnamefont {Adiyatullin}}, \emph {et~al.},\ }\href {https://doi.org/10.1038/s41567-022-01653-5} {\bibfield  {journal} {\bibinfo  {journal} {Nature Physics}\ }\textbf {\bibinfo {volume} {18}},\ \bibinfo {pages} {925} (\bibinfo {year} {2022})}\BibitemShut {NoStop}%
\bibitem [{\citenamefont {Hines}\ \emph {et~al.}(2023)\citenamefont {Hines}, \citenamefont {Rajagopal}, \citenamefont {Moreau}, \citenamefont {Wahrman}, \citenamefont {Lewis}, \citenamefont {Markovi{\'c}},\ and\ \citenamefont {Schleier-Smith}}]{hines2023}%
  \BibitemOpen
  \bibfield  {author} {\bibinfo {author} {\bibfnamefont {J.~A.}\ \bibnamefont {Hines}}, \bibinfo {author} {\bibfnamefont {S.~V.}\ \bibnamefont {Rajagopal}}, \bibinfo {author} {\bibfnamefont {G.~L.}\ \bibnamefont {Moreau}}, \bibinfo {author} {\bibfnamefont {M.~D.}\ \bibnamefont {Wahrman}}, \bibinfo {author} {\bibfnamefont {N.~A.}\ \bibnamefont {Lewis}}, \bibinfo {author} {\bibfnamefont {O.}~\bibnamefont {Markovi{\'c}}},\ and\ \bibinfo {author} {\bibfnamefont {M.}~\bibnamefont {Schleier-Smith}},\ }\href@noop {} {\bibfield  {journal} {\bibinfo  {journal} {Physical Review Letters}\ }\textbf {\bibinfo {volume} {131}},\ \bibinfo {pages} {063401} (\bibinfo {year} {2023})}\BibitemShut {NoStop}%
\bibitem [{\citenamefont {Bluvstein}\ \emph {et~al.}(2023)\citenamefont {Bluvstein}, \citenamefont {Evered}, \citenamefont {Geim}, \citenamefont {Li}, \citenamefont {Zhou}, \citenamefont {Manovitz}, \citenamefont {Ebadi}, \citenamefont {Cain}, \citenamefont {Kalinowski}, \citenamefont {Hangleiter}, \citenamefont {Bonilla~Ataides}, \citenamefont {Maskara}, \citenamefont {Cong}, \citenamefont {Gao}, \citenamefont {Sales~Rodriguez}, \citenamefont {Karolyshyn}, \citenamefont {Semeghini}, \citenamefont {Gullans}, \citenamefont {Greiner}, \citenamefont {Vuletić},\ and\ \citenamefont {Lukin}}]{Bluvstein_2023}%
  \BibitemOpen
  \bibfield  {author} {\bibinfo {author} {\bibfnamefont {D.}~\bibnamefont {Bluvstein}}, \bibinfo {author} {\bibfnamefont {S.~J.}\ \bibnamefont {Evered}}, \bibinfo {author} {\bibfnamefont {A.~A.}\ \bibnamefont {Geim}}, \bibinfo {author} {\bibfnamefont {S.~H.}\ \bibnamefont {Li}}, \bibinfo {author} {\bibfnamefont {H.}~\bibnamefont {Zhou}}, \bibinfo {author} {\bibfnamefont {T.}~\bibnamefont {Manovitz}}, \bibinfo {author} {\bibfnamefont {S.}~\bibnamefont {Ebadi}}, \bibinfo {author} {\bibfnamefont {M.}~\bibnamefont {Cain}}, \bibinfo {author} {\bibfnamefont {M.}~\bibnamefont {Kalinowski}}, \bibinfo {author} {\bibfnamefont {D.}~\bibnamefont {Hangleiter}}, \bibinfo {author} {\bibfnamefont {J.~P.}\ \bibnamefont {Bonilla~Ataides}}, \bibinfo {author} {\bibfnamefont {N.}~\bibnamefont {Maskara}}, \bibinfo {author} {\bibfnamefont {I.}~\bibnamefont {Cong}}, \bibinfo {author} {\bibfnamefont {X.}~\bibnamefont {Gao}}, \bibinfo {author} {\bibfnamefont {P.}~\bibnamefont {Sales~Rodriguez}}, \bibinfo {author} {\bibfnamefont {T.}~\bibnamefont {Karolyshyn}}, \bibinfo {author} {\bibfnamefont {G.}~\bibnamefont {Semeghini}}, \bibinfo {author} {\bibfnamefont {M.~J.}\ \bibnamefont {Gullans}}, \bibinfo {author} {\bibfnamefont {M.}~\bibnamefont {Greiner}}, \bibinfo {author} {\bibfnamefont {V.}~\bibnamefont {Vuletić}},\ and\ \bibinfo {author} {\bibfnamefont {M.~D.}\ \bibnamefont {Lukin}},\ }\href {https://doi.org/10.1038/s41586-023-06927-3} {\bibfield  {journal} {\bibinfo  {journal} {Nature}\ }\textbf {\bibinfo {volume} {626}},\ \bibinfo {pages} {58–65} (\bibinfo {year} {2023})}\BibitemShut {NoStop}%
\bibitem [{\citenamefont {Bluvstein}\ \emph {et~al.}(2022)\citenamefont {Bluvstein}, \citenamefont {Levine}, \citenamefont {Semeghini}, \citenamefont {Wang}, \citenamefont {Ebadi}, \citenamefont {Kalinowski}, \citenamefont {Keesling}, \citenamefont {Maskara}, \citenamefont {Pichler}, \citenamefont {Greiner}, \citenamefont {Vuletić},\ and\ \citenamefont {Lukin}}]{Bluvstein_2022}%
  \BibitemOpen
  \bibfield  {author} {\bibinfo {author} {\bibfnamefont {D.}~\bibnamefont {Bluvstein}}, \bibinfo {author} {\bibfnamefont {H.}~\bibnamefont {Levine}}, \bibinfo {author} {\bibfnamefont {G.}~\bibnamefont {Semeghini}}, \bibinfo {author} {\bibfnamefont {T.~T.}\ \bibnamefont {Wang}}, \bibinfo {author} {\bibfnamefont {S.}~\bibnamefont {Ebadi}}, \bibinfo {author} {\bibfnamefont {M.}~\bibnamefont {Kalinowski}}, \bibinfo {author} {\bibfnamefont {A.}~\bibnamefont {Keesling}}, \bibinfo {author} {\bibfnamefont {N.}~\bibnamefont {Maskara}}, \bibinfo {author} {\bibfnamefont {H.}~\bibnamefont {Pichler}}, \bibinfo {author} {\bibfnamefont {M.}~\bibnamefont {Greiner}}, \bibinfo {author} {\bibfnamefont {V.}~\bibnamefont {Vuletić}},\ and\ \bibinfo {author} {\bibfnamefont {M.~D.}\ \bibnamefont {Lukin}},\ }\href {https://doi.org/10.1038/s41586-022-04592-6} {\bibfield  {journal} {\bibinfo  {journal} {Nature}\ }\textbf {\bibinfo {volume} {604}},\ \bibinfo {pages} {451–456} (\bibinfo {year} {2022})}\BibitemShut {NoStop}%
\bibitem [{\citenamefont {Xu}\ \emph {et~al.}(2024)\citenamefont {Xu}, \citenamefont {Bonilla~Ataides}, \citenamefont {Pattison} \emph {et~al.}}]{Xu2024}%
  \BibitemOpen
  \bibfield  {author} {\bibinfo {author} {\bibfnamefont {Q.}~\bibnamefont {Xu}}, \bibinfo {author} {\bibfnamefont {J.~P.}\ \bibnamefont {Bonilla~Ataides}}, \bibinfo {author} {\bibfnamefont {C.~A.}\ \bibnamefont {Pattison}}, \emph {et~al.},\ }\href {https://doi.org/10.1038/s41567-024-02479-z} {\bibfield  {journal} {\bibinfo  {journal} {Nature Physics}\ }\textbf {\bibinfo {volume} {20}},\ \bibinfo {pages} {1084} (\bibinfo {year} {2024})}\BibitemShut {NoStop}%
\bibitem [{\citenamefont {Bornet}\ \emph {et~al.}(2023{\natexlab{a}})\citenamefont {Bornet}, \citenamefont {Emperauger}, \citenamefont {Chen}, \citenamefont {Ye}, \citenamefont {Block}, \citenamefont {Bintz}, \citenamefont {Boyd}, \citenamefont {Barredo}, \citenamefont {Comparin}, \citenamefont {Mezzacapo}, \citenamefont {Roscilde}, \citenamefont {Lahaye}, \citenamefont {Yao},\ and\ \citenamefont {Browaeys}}]{Bornet_2023}%
  \BibitemOpen
  \bibfield  {author} {\bibinfo {author} {\bibfnamefont {G.}~\bibnamefont {Bornet}}, \bibinfo {author} {\bibfnamefont {G.}~\bibnamefont {Emperauger}}, \bibinfo {author} {\bibfnamefont {C.}~\bibnamefont {Chen}}, \bibinfo {author} {\bibfnamefont {B.}~\bibnamefont {Ye}}, \bibinfo {author} {\bibfnamefont {M.}~\bibnamefont {Block}}, \bibinfo {author} {\bibfnamefont {M.}~\bibnamefont {Bintz}}, \bibinfo {author} {\bibfnamefont {J.~A.}\ \bibnamefont {Boyd}}, \bibinfo {author} {\bibfnamefont {D.}~\bibnamefont {Barredo}}, \bibinfo {author} {\bibfnamefont {T.}~\bibnamefont {Comparin}}, \bibinfo {author} {\bibfnamefont {F.}~\bibnamefont {Mezzacapo}}, \bibinfo {author} {\bibfnamefont {T.}~\bibnamefont {Roscilde}}, \bibinfo {author} {\bibfnamefont {T.}~\bibnamefont {Lahaye}}, \bibinfo {author} {\bibfnamefont {N.~Y.}\ \bibnamefont {Yao}},\ and\ \bibinfo {author} {\bibfnamefont {A.}~\bibnamefont {Browaeys}},\ }\href {https://doi.org/10.1038/s41586-023-06414-9} {\bibfield  {journal} {\bibinfo  {journal} {Nature}\ }\textbf {\bibinfo {volume} {621}},\ \bibinfo {pages} {728–733} (\bibinfo {year} {2023}{\natexlab{a}})}\BibitemShut {NoStop}%
\bibitem [{\citenamefont {Eckner}\ \emph {et~al.}(2023)\citenamefont {Eckner}, \citenamefont {Darkwah~Oppong}, \citenamefont {Cao}, \citenamefont {Young}, \citenamefont {Milner}, \citenamefont {Robinson}, \citenamefont {Ye},\ and\ \citenamefont {Kaufman}}]{eckner2023}%
  \BibitemOpen
  \bibfield  {author} {\bibinfo {author} {\bibfnamefont {W.~J.}\ \bibnamefont {Eckner}}, \bibinfo {author} {\bibfnamefont {N.}~\bibnamefont {Darkwah~Oppong}}, \bibinfo {author} {\bibfnamefont {A.}~\bibnamefont {Cao}}, \bibinfo {author} {\bibfnamefont {A.~W.}\ \bibnamefont {Young}}, \bibinfo {author} {\bibfnamefont {W.~R.}\ \bibnamefont {Milner}}, \bibinfo {author} {\bibfnamefont {J.~M.}\ \bibnamefont {Robinson}}, \bibinfo {author} {\bibfnamefont {J.}~\bibnamefont {Ye}},\ and\ \bibinfo {author} {\bibfnamefont {A.~M.}\ \bibnamefont {Kaufman}},\ }\href@noop {} {\bibfield  {journal} {\bibinfo  {journal} {Nature}\ }\textbf {\bibinfo {volume} {621}},\ \bibinfo {pages} {734} (\bibinfo {year} {2023})}\BibitemShut {NoStop}%
\bibitem [{\citenamefont {Brennen}\ \emph {et~al.}(1999)\citenamefont {Brennen}, \citenamefont {Caves}, \citenamefont {Jessen},\ and\ \citenamefont {Deutsch}}]{Brennen_1999_QL}%
  \BibitemOpen
  \bibfield  {author} {\bibinfo {author} {\bibfnamefont {G.~K.}\ \bibnamefont {Brennen}}, \bibinfo {author} {\bibfnamefont {C.~M.}\ \bibnamefont {Caves}}, \bibinfo {author} {\bibfnamefont {P.~S.}\ \bibnamefont {Jessen}},\ and\ \bibinfo {author} {\bibfnamefont {I.~H.}\ \bibnamefont {Deutsch}},\ }\href {https://doi.org/10.1103/PhysRevLett.82.1060} {\bibfield  {journal} {\bibinfo  {journal} {Phys. Rev. Lett.}\ }\textbf {\bibinfo {volume} {82}},\ \bibinfo {pages} {1060} (\bibinfo {year} {1999})}\BibitemShut {NoStop}%
\bibitem [{\citenamefont {Jaksch}\ \emph {et~al.}(2000)\citenamefont {Jaksch}, \citenamefont {Cirac}, \citenamefont {Zoller}, \citenamefont {Rolston}, \citenamefont {C\^ot\'e},\ and\ \citenamefont {Lukin}}]{Jaksch_2000_Fast_quantum}%
  \BibitemOpen
  \bibfield  {author} {\bibinfo {author} {\bibfnamefont {D.}~\bibnamefont {Jaksch}}, \bibinfo {author} {\bibfnamefont {J.~I.}\ \bibnamefont {Cirac}}, \bibinfo {author} {\bibfnamefont {P.}~\bibnamefont {Zoller}}, \bibinfo {author} {\bibfnamefont {S.~L.}\ \bibnamefont {Rolston}}, \bibinfo {author} {\bibfnamefont {R.}~\bibnamefont {C\^ot\'e}},\ and\ \bibinfo {author} {\bibfnamefont {M.~D.}\ \bibnamefont {Lukin}},\ }\href {https://doi.org/10.1103/PhysRevLett.85.2208} {\bibfield  {journal} {\bibinfo  {journal} {Phys. Rev. Lett.}\ }\textbf {\bibinfo {volume} {85}},\ \bibinfo {pages} {2208} (\bibinfo {year} {2000})}\BibitemShut {NoStop}%
\bibitem [{\citenamefont {Brion}\ \emph {et~al.}(2007)\citenamefont {Brion}, \citenamefont {M\o{}lmer},\ and\ \citenamefont {Saffman}}]{Brion_2007_QCE}%
  \BibitemOpen
  \bibfield  {author} {\bibinfo {author} {\bibfnamefont {E.}~\bibnamefont {Brion}}, \bibinfo {author} {\bibfnamefont {K.}~\bibnamefont {M\o{}lmer}},\ and\ \bibinfo {author} {\bibfnamefont {M.}~\bibnamefont {Saffman}},\ }\href {https://doi.org/10.1103/PhysRevLett.99.260501} {\bibfield  {journal} {\bibinfo  {journal} {Phys. Rev. Lett.}\ }\textbf {\bibinfo {volume} {99}},\ \bibinfo {pages} {260501} (\bibinfo {year} {2007})}\BibitemShut {NoStop}%
\bibitem [{\citenamefont {Mølmer}\ \emph {et~al.}(2011)\citenamefont {Mølmer}, \citenamefont {Isenhower},\ and\ \citenamefont {Saffman}}]{Mølmer_2011}%
  \BibitemOpen
  \bibfield  {author} {\bibinfo {author} {\bibfnamefont {K.}~\bibnamefont {Mølmer}}, \bibinfo {author} {\bibfnamefont {L.}~\bibnamefont {Isenhower}},\ and\ \bibinfo {author} {\bibfnamefont {M.}~\bibnamefont {Saffman}},\ }\href {https://doi.org/10.1088/0953-4075/44/18/184016} {\bibfield  {journal} {\bibinfo  {journal} {Journal of Physics B: Atomic, Molecular and Optical Physics}\ }\textbf {\bibinfo {volume} {44}},\ \bibinfo {pages} {184016} (\bibinfo {year} {2011})}\BibitemShut {NoStop}%
\bibitem [{\citenamefont {Saffman}(2016)}]{Saffman_2016}%
  \BibitemOpen
  \bibfield  {author} {\bibinfo {author} {\bibfnamefont {M.}~\bibnamefont {Saffman}},\ }\href {https://doi.org/10.1088/0953-4075/49/20/202001} {\bibfield  {journal} {\bibinfo  {journal} {Journal of Physics B: Atomic, Molecular and Optical Physics}\ }\textbf {\bibinfo {volume} {49}},\ \bibinfo {pages} {202001} (\bibinfo {year} {2016})}\BibitemShut {NoStop}%
\bibitem [{\citenamefont {Weiss}\ and\ \citenamefont {Saffman}(2017)}]{Weiss_2017}%
  \BibitemOpen
  \bibfield  {author} {\bibinfo {author} {\bibfnamefont {D.~S.}\ \bibnamefont {Weiss}}\ and\ \bibinfo {author} {\bibfnamefont {M.}~\bibnamefont {Saffman}},\ }\href {https://doi.org/10.1063/PT.3.3626} {\bibfield  {journal} {\bibinfo  {journal} {Physics Today}\ }\textbf {\bibinfo {volume} {70}},\ \bibinfo {pages} {44} (\bibinfo {year} {2017})}\BibitemShut {NoStop}%
\bibitem [{\citenamefont {Adams}\ \emph {et~al.}(2019)\citenamefont {Adams}, \citenamefont {Pritchard},\ and\ \citenamefont {Shaffer}}]{Adams_2020}%
  \BibitemOpen
  \bibfield  {author} {\bibinfo {author} {\bibfnamefont {C.~S.}\ \bibnamefont {Adams}}, \bibinfo {author} {\bibfnamefont {J.~D.}\ \bibnamefont {Pritchard}},\ and\ \bibinfo {author} {\bibfnamefont {J.~P.}\ \bibnamefont {Shaffer}},\ }\href {https://doi.org/10.1088/1361-6455/ab52ef} {\bibfield  {journal} {\bibinfo  {journal} {Journal of Physics B: Atomic, Molecular and Optical Physics}\ }\textbf {\bibinfo {volume} {53}},\ \bibinfo {pages} {012002} (\bibinfo {year} {2019})}\BibitemShut {NoStop}%
\bibitem [{\citenamefont {Henriet}\ \emph {et~al.}(2020)\citenamefont {Henriet}, \citenamefont {Beguin}, \citenamefont {Signoles}, \citenamefont {Lahaye}, \citenamefont {Browaeys}, \citenamefont {Reymond},\ and\ \citenamefont {Jurczak}}]{Henriet2020quantumcomputing}%
  \BibitemOpen
  \bibfield  {author} {\bibinfo {author} {\bibfnamefont {L.}~\bibnamefont {Henriet}}, \bibinfo {author} {\bibfnamefont {L.}~\bibnamefont {Beguin}}, \bibinfo {author} {\bibfnamefont {A.}~\bibnamefont {Signoles}}, \bibinfo {author} {\bibfnamefont {T.}~\bibnamefont {Lahaye}}, \bibinfo {author} {\bibfnamefont {A.}~\bibnamefont {Browaeys}}, \bibinfo {author} {\bibfnamefont {G.-O.}\ \bibnamefont {Reymond}},\ and\ \bibinfo {author} {\bibfnamefont {C.}~\bibnamefont {Jurczak}},\ }\href {https://doi.org/10.22331/q-2020-09-21-327} {\bibfield  {journal} {\bibinfo  {journal} {{Quantum}}\ }\textbf {\bibinfo {volume} {4}},\ \bibinfo {pages} {327} (\bibinfo {year} {2020})}\BibitemShut {NoStop}%
\bibitem [{\citenamefont {Levine}\ \emph {et~al.}(2019)\citenamefont {Levine}, \citenamefont {Keesling}, \citenamefont {Omran}, \citenamefont {Pichler}, \citenamefont {Choi}, \citenamefont {Samajdar}, \citenamefont {Schkolnik}, \citenamefont {Wang}, \citenamefont {Zibrov}, \citenamefont {Endres}, \citenamefont {Greiner}, \citenamefont {Vuleti{\'c}},\ and\ \citenamefont {Lukin}}]{levine2019parallel}%
  \BibitemOpen
  \bibfield  {author} {\bibinfo {author} {\bibfnamefont {H.}~\bibnamefont {Levine}}, \bibinfo {author} {\bibfnamefont {A.}~\bibnamefont {Keesling}}, \bibinfo {author} {\bibfnamefont {A.}~\bibnamefont {Omran}}, \bibinfo {author} {\bibfnamefont {H.}~\bibnamefont {Pichler}}, \bibinfo {author} {\bibfnamefont {S.}~\bibnamefont {Choi}}, \bibinfo {author} {\bibfnamefont {R.}~\bibnamefont {Samajdar}}, \bibinfo {author} {\bibfnamefont {V.}~\bibnamefont {Schkolnik}}, \bibinfo {author} {\bibfnamefont {H.}~\bibnamefont {Wang}}, \bibinfo {author} {\bibfnamefont {A.~S.}\ \bibnamefont {Zibrov}}, \bibinfo {author} {\bibfnamefont {M.}~\bibnamefont {Endres}}, \bibinfo {author} {\bibfnamefont {M.}~\bibnamefont {Greiner}}, \bibinfo {author} {\bibfnamefont {V.}~\bibnamefont {Vuleti{\'c}}},\ and\ \bibinfo {author} {\bibfnamefont {M.~D.}\ \bibnamefont {Lukin}},\ }\href {https://doi.org/10.1103/PhysRevLett.123.170503} {\bibfield  {journal} {\bibinfo  {journal} {Physical Review Letters}\ }\textbf {\bibinfo {volume} {123}},\ \bibinfo {pages} {170503} (\bibinfo {year} {2019})}\BibitemShut {NoStop}%
\bibitem [{\citenamefont {Cong}\ \emph {et~al.}(2022)\citenamefont {Cong}, \citenamefont {Levine}, \citenamefont {Keesling}, \citenamefont {Bluvstein}, \citenamefont {Wang},\ and\ \citenamefont {Lukin}}]{CongHardware2022}%
  \BibitemOpen
  \bibfield  {author} {\bibinfo {author} {\bibfnamefont {I.}~\bibnamefont {Cong}}, \bibinfo {author} {\bibfnamefont {H.}~\bibnamefont {Levine}}, \bibinfo {author} {\bibfnamefont {A.}~\bibnamefont {Keesling}}, \bibinfo {author} {\bibfnamefont {D.}~\bibnamefont {Bluvstein}}, \bibinfo {author} {\bibfnamefont {S.-T.}\ \bibnamefont {Wang}},\ and\ \bibinfo {author} {\bibfnamefont {M.~D.}\ \bibnamefont {Lukin}},\ }\href {https://doi.org/10.1103/PhysRevX.12.021049} {\bibfield  {journal} {\bibinfo  {journal} {Phys. Rev. X}\ }\textbf {\bibinfo {volume} {12}},\ \bibinfo {pages} {021049} (\bibinfo {year} {2022})}\BibitemShut {NoStop}%
\bibitem [{\citenamefont {Wu}\ \emph {et~al.}(2022)\citenamefont {Wu}, \citenamefont {Kolkowitz}, \citenamefont {Puri} \emph {et~al.}}]{wu2022erasure}%
  \BibitemOpen
  \bibfield  {author} {\bibinfo {author} {\bibfnamefont {Y.}~\bibnamefont {Wu}}, \bibinfo {author} {\bibfnamefont {S.}~\bibnamefont {Kolkowitz}}, \bibinfo {author} {\bibfnamefont {S.}~\bibnamefont {Puri}}, \emph {et~al.},\ }\href {https://doi.org/10.1038/s41467-022-32094-6} {\bibfield  {journal} {\bibinfo  {journal} {Nature Communications}\ }\textbf {\bibinfo {volume} {13}},\ \bibinfo {pages} {4657} (\bibinfo {year} {2022})}\BibitemShut {NoStop}%
\bibitem [{\citenamefont {Auger}\ \emph {et~al.}(2017)\citenamefont {Auger}, \citenamefont {Bergamini},\ and\ \citenamefont {Browne}}]{AugerBlueprint2017}%
  \BibitemOpen
  \bibfield  {author} {\bibinfo {author} {\bibfnamefont {J.~M.}\ \bibnamefont {Auger}}, \bibinfo {author} {\bibfnamefont {S.}~\bibnamefont {Bergamini}},\ and\ \bibinfo {author} {\bibfnamefont {D.~E.}\ \bibnamefont {Browne}},\ }\href {https://doi.org/10.1103/PhysRevA.96.052320} {\bibfield  {journal} {\bibinfo  {journal} {Phys. Rev. A}\ }\textbf {\bibinfo {volume} {96}},\ \bibinfo {pages} {052320} (\bibinfo {year} {2017})}\BibitemShut {NoStop}%
\bibitem [{\citenamefont {Evered}\ \emph {et~al.}(2023)\citenamefont {Evered}, \citenamefont {Bluvstein}, \citenamefont {Kalinowski} \emph {et~al.}}]{Evered2023}%
  \BibitemOpen
  \bibfield  {author} {\bibinfo {author} {\bibfnamefont {S.}~\bibnamefont {Evered}}, \bibinfo {author} {\bibfnamefont {D.}~\bibnamefont {Bluvstein}}, \bibinfo {author} {\bibfnamefont {M.}~\bibnamefont {Kalinowski}}, \emph {et~al.},\ }\href {https://doi.org/10.1038/s41586-023-06481-y} {\bibfield  {journal} {\bibinfo  {journal} {Nature}\ }\textbf {\bibinfo {volume} {622}},\ \bibinfo {pages} {268} (\bibinfo {year} {2023})}\BibitemShut {NoStop}%
\bibitem [{\citenamefont {Nikolov}\ \emph {et~al.}(2023)\citenamefont {Nikolov}, \citenamefont {Diamond-Hitchcock}, \citenamefont {Bass}, \citenamefont {Spong},\ and\ \citenamefont {Pritchard}}]{nikolov2023}%
  \BibitemOpen
  \bibfield  {author} {\bibinfo {author} {\bibfnamefont {B.}~\bibnamefont {Nikolov}}, \bibinfo {author} {\bibfnamefont {E.}~\bibnamefont {Diamond-Hitchcock}}, \bibinfo {author} {\bibfnamefont {J.}~\bibnamefont {Bass}}, \bibinfo {author} {\bibfnamefont {N.}~\bibnamefont {Spong}},\ and\ \bibinfo {author} {\bibfnamefont {J.}~\bibnamefont {Pritchard}},\ }\href@noop {} {\bibfield  {journal} {\bibinfo  {journal} {Physical Review Letters}\ }\textbf {\bibinfo {volume} {131}},\ \bibinfo {pages} {030602} (\bibinfo {year} {2023})}\BibitemShut {NoStop}%
\bibitem [{\citenamefont {Pelegr}\ \emph {et~al.}(2022)\citenamefont {Pelegr}, \citenamefont {Daley},\ and\ \citenamefont {Pritchard}}]{Pelegri_2022}%
  \BibitemOpen
  \bibfield  {author} {\bibinfo {author} {\bibfnamefont {G.}~\bibnamefont {Pelegr}}, \bibinfo {author} {\bibfnamefont {A.~J.}\ \bibnamefont {Daley}},\ and\ \bibinfo {author} {\bibfnamefont {J.~D.}\ \bibnamefont {Pritchard}},\ }\href {https://doi.org/10.1088/2058-9565/ac823a} {\bibfield  {journal} {\bibinfo  {journal} {Quantum Science and Technology}\ }\textbf {\bibinfo {volume} {7}},\ \bibinfo {pages} {045020} (\bibinfo {year} {2022})}\BibitemShut {NoStop}%
\bibitem [{\citenamefont {Jandura}\ and\ \citenamefont {Pupillo}(2022)}]{Jandura_2022}%
  \BibitemOpen
  \bibfield  {author} {\bibinfo {author} {\bibfnamefont {S.}~\bibnamefont {Jandura}}\ and\ \bibinfo {author} {\bibfnamefont {G.}~\bibnamefont {Pupillo}},\ }\href {https://doi.org/10.22331/q-2022-05-13-712} {\bibfield  {journal} {\bibinfo  {journal} {Quantum}\ }\textbf {\bibinfo {volume} {6}},\ \bibinfo {pages} {712} (\bibinfo {year} {2022})}\BibitemShut {NoStop}%
\bibitem [{\citenamefont {Pagano}\ \emph {et~al.}(2022)\citenamefont {Pagano}, \citenamefont {Weber}, \citenamefont {Jaschke}, \citenamefont {Pfau}, \citenamefont {Meinert}, \citenamefont {Montangero},\ and\ \citenamefont {B\"uchler}}]{Pagano_2022}%
  \BibitemOpen
  \bibfield  {author} {\bibinfo {author} {\bibfnamefont {A.}~\bibnamefont {Pagano}}, \bibinfo {author} {\bibfnamefont {S.}~\bibnamefont {Weber}}, \bibinfo {author} {\bibfnamefont {D.}~\bibnamefont {Jaschke}}, \bibinfo {author} {\bibfnamefont {T.}~\bibnamefont {Pfau}}, \bibinfo {author} {\bibfnamefont {F.}~\bibnamefont {Meinert}}, \bibinfo {author} {\bibfnamefont {S.}~\bibnamefont {Montangero}},\ and\ \bibinfo {author} {\bibfnamefont {H.~P.}\ \bibnamefont {B\"uchler}},\ }\href {https://doi.org/10.1103/PhysRevResearch.4.033019} {\bibfield  {journal} {\bibinfo  {journal} {Phys. Rev. Res.}\ }\textbf {\bibinfo {volume} {4}},\ \bibinfo {pages} {033019} (\bibinfo {year} {2022})}\BibitemShut {NoStop}%
\bibitem [{\citenamefont {Fromonteil}\ \emph {et~al.}(2024)\citenamefont {Fromonteil}, \citenamefont {Tricarico}, \citenamefont {Cesa},\ and\ \citenamefont {Pichler}}]{fromonteil2024}%
  \BibitemOpen
  \bibfield  {author} {\bibinfo {author} {\bibfnamefont {C.}~\bibnamefont {Fromonteil}}, \bibinfo {author} {\bibfnamefont {R.}~\bibnamefont {Tricarico}}, \bibinfo {author} {\bibfnamefont {F.}~\bibnamefont {Cesa}},\ and\ \bibinfo {author} {\bibfnamefont {H.}~\bibnamefont {Pichler}},\ }\href@noop {} {\bibfield  {journal} {\bibinfo  {journal} {Physical Review Research}\ }\textbf {\bibinfo {volume} {6}},\ \bibinfo {pages} {033333} (\bibinfo {year} {2024})}\BibitemShut {NoStop}%
\bibitem [{\citenamefont {Bernien}\ \emph {et~al.}(2017)\citenamefont {Bernien}, \citenamefont {Schwartz}, \citenamefont {Keesling}, \citenamefont {Levine}, \citenamefont {Omran}, \citenamefont {Pichler}, \citenamefont {Choi}, \citenamefont {Zibrov}, \citenamefont {Endres}, \citenamefont {Greiner}, \citenamefont {Vuletić},\ and\ \citenamefont {Lukin}}]{Bernien_2017}%
  \BibitemOpen
  \bibfield  {author} {\bibinfo {author} {\bibfnamefont {H.}~\bibnamefont {Bernien}}, \bibinfo {author} {\bibfnamefont {S.}~\bibnamefont {Schwartz}}, \bibinfo {author} {\bibfnamefont {A.}~\bibnamefont {Keesling}}, \bibinfo {author} {\bibfnamefont {H.}~\bibnamefont {Levine}}, \bibinfo {author} {\bibfnamefont {A.}~\bibnamefont {Omran}}, \bibinfo {author} {\bibfnamefont {H.}~\bibnamefont {Pichler}}, \bibinfo {author} {\bibfnamefont {S.}~\bibnamefont {Choi}}, \bibinfo {author} {\bibfnamefont {A.~S.}\ \bibnamefont {Zibrov}}, \bibinfo {author} {\bibfnamefont {M.}~\bibnamefont {Endres}}, \bibinfo {author} {\bibfnamefont {M.}~\bibnamefont {Greiner}}, \bibinfo {author} {\bibfnamefont {V.}~\bibnamefont {Vuletić}},\ and\ \bibinfo {author} {\bibfnamefont {M.~D.}\ \bibnamefont {Lukin}},\ }\href {https://doi.org/10.1038/nature24622} {\bibfield  {journal} {\bibinfo  {journal} {Nature}\ }\textbf {\bibinfo {volume} {551}},\ \bibinfo {pages} {579–584} (\bibinfo {year} {2017})}\BibitemShut {NoStop}%
\bibitem [{\citenamefont {Browaeys}\ and\ \citenamefont {Lahaye}(2020)}]{Browaeys_2020}%
  \BibitemOpen
  \bibfield  {author} {\bibinfo {author} {\bibfnamefont {A.}~\bibnamefont {Browaeys}}\ and\ \bibinfo {author} {\bibfnamefont {T.}~\bibnamefont {Lahaye}},\ }\href {https://doi.org/10.1038/s41567-019-0733-z} {\bibfield  {journal} {\bibinfo  {journal} {Nature Physics}\ }\textbf {\bibinfo {volume} {16}},\ \bibinfo {pages} {132–142} (\bibinfo {year} {2020})}\BibitemShut {NoStop}%
\bibitem [{\citenamefont {Omran}\ \emph {et~al.}(2019)\citenamefont {Omran}, \citenamefont {Levine}, \citenamefont {Keesling}, \citenamefont {Semeghini}, \citenamefont {Wang}, \citenamefont {Ebadi}, \citenamefont {Bernien}, \citenamefont {Zibrov}, \citenamefont {Pichler}, \citenamefont {Choi}, \citenamefont {Cui}, \citenamefont {Rossignolo}, \citenamefont {Rembold}, \citenamefont {Montangero}, \citenamefont {Calarco}, \citenamefont {Endres}, \citenamefont {Greiner}, \citenamefont {Vuletić},\ and\ \citenamefont {Lukin}}]{Omran_2019}%
  \BibitemOpen
  \bibfield  {author} {\bibinfo {author} {\bibfnamefont {A.}~\bibnamefont {Omran}}, \bibinfo {author} {\bibfnamefont {H.}~\bibnamefont {Levine}}, \bibinfo {author} {\bibfnamefont {A.}~\bibnamefont {Keesling}}, \bibinfo {author} {\bibfnamefont {G.}~\bibnamefont {Semeghini}}, \bibinfo {author} {\bibfnamefont {T.~T.}\ \bibnamefont {Wang}}, \bibinfo {author} {\bibfnamefont {S.}~\bibnamefont {Ebadi}}, \bibinfo {author} {\bibfnamefont {H.}~\bibnamefont {Bernien}}, \bibinfo {author} {\bibfnamefont {A.~S.}\ \bibnamefont {Zibrov}}, \bibinfo {author} {\bibfnamefont {H.}~\bibnamefont {Pichler}}, \bibinfo {author} {\bibfnamefont {S.}~\bibnamefont {Choi}}, \bibinfo {author} {\bibfnamefont {J.}~\bibnamefont {Cui}}, \bibinfo {author} {\bibfnamefont {M.}~\bibnamefont {Rossignolo}}, \bibinfo {author} {\bibfnamefont {P.}~\bibnamefont {Rembold}}, \bibinfo {author} {\bibfnamefont {S.}~\bibnamefont {Montangero}}, \bibinfo {author} {\bibfnamefont {T.}~\bibnamefont {Calarco}}, \bibinfo {author} {\bibfnamefont {M.}~\bibnamefont {Endres}}, \bibinfo {author} {\bibfnamefont {M.}~\bibnamefont {Greiner}}, \bibinfo {author} {\bibfnamefont {V.}~\bibnamefont {Vuletić}},\ and\ \bibinfo {author} {\bibfnamefont {M.~D.}\ \bibnamefont {Lukin}},\ }\href {https://doi.org/10.1126/science.aax9743} {\bibfield  {journal} {\bibinfo  {journal} {Science}\ }\textbf {\bibinfo {volume} {365}},\ \bibinfo {pages} {570} (\bibinfo {year} {2019})}\BibitemShut {NoStop}%
\bibitem [{\citenamefont {Verresen}\ \emph {et~al.}(2021)\citenamefont {Verresen}, \citenamefont {Lukin},\ and\ \citenamefont {Vishwanath}}]{Veressen_2021}%
  \BibitemOpen
  \bibfield  {author} {\bibinfo {author} {\bibfnamefont {R.}~\bibnamefont {Verresen}}, \bibinfo {author} {\bibfnamefont {M.~D.}\ \bibnamefont {Lukin}},\ and\ \bibinfo {author} {\bibfnamefont {A.}~\bibnamefont {Vishwanath}},\ }\href {https://doi.org/10.1103/PhysRevX.11.031005} {\bibfield  {journal} {\bibinfo  {journal} {Phys. Rev. X}\ }\textbf {\bibinfo {volume} {11}},\ \bibinfo {pages} {031005} (\bibinfo {year} {2021})}\BibitemShut {NoStop}%
\bibitem [{\citenamefont {Semeghini}\ \emph {et~al.}(2021)\citenamefont {Semeghini}, \citenamefont {Levine}, \citenamefont {Keesling}, \citenamefont {Ebadi}, \citenamefont {Wang}, \citenamefont {Bluvstein}, \citenamefont {Verresen}, \citenamefont {Pichler}, \citenamefont {Kalinowski}, \citenamefont {Samajdar}, \citenamefont {Omran}, \citenamefont {Sachdev}, \citenamefont {Vishwanath}, \citenamefont {Greiner}, \citenamefont {Vuletić},\ and\ \citenamefont {Lukin}}]{Semeghini_2021}%
  \BibitemOpen
  \bibfield  {author} {\bibinfo {author} {\bibfnamefont {G.}~\bibnamefont {Semeghini}}, \bibinfo {author} {\bibfnamefont {H.}~\bibnamefont {Levine}}, \bibinfo {author} {\bibfnamefont {A.}~\bibnamefont {Keesling}}, \bibinfo {author} {\bibfnamefont {S.}~\bibnamefont {Ebadi}}, \bibinfo {author} {\bibfnamefont {T.~T.}\ \bibnamefont {Wang}}, \bibinfo {author} {\bibfnamefont {D.}~\bibnamefont {Bluvstein}}, \bibinfo {author} {\bibfnamefont {R.}~\bibnamefont {Verresen}}, \bibinfo {author} {\bibfnamefont {H.}~\bibnamefont {Pichler}}, \bibinfo {author} {\bibfnamefont {M.}~\bibnamefont {Kalinowski}}, \bibinfo {author} {\bibfnamefont {R.}~\bibnamefont {Samajdar}}, \bibinfo {author} {\bibfnamefont {A.}~\bibnamefont {Omran}}, \bibinfo {author} {\bibfnamefont {S.}~\bibnamefont {Sachdev}}, \bibinfo {author} {\bibfnamefont {A.}~\bibnamefont {Vishwanath}}, \bibinfo {author} {\bibfnamefont {M.}~\bibnamefont {Greiner}}, \bibinfo {author} {\bibfnamefont {V.}~\bibnamefont {Vuletić}},\ and\ \bibinfo {author} {\bibfnamefont {M.~D.}\ \bibnamefont {Lukin}},\ }\href {https://doi.org/10.1126/science.abi8794} {\bibfield  {journal} {\bibinfo  {journal} {Science}\ }\textbf {\bibinfo {volume} {374}},\ \bibinfo {pages} {1242} (\bibinfo {year} {2021})}\BibitemShut {NoStop}%
\bibitem [{\citenamefont {Labuhn}\ \emph {et~al.}(2016)\citenamefont {Labuhn}, \citenamefont {Barredo}, \citenamefont {Ravets}, \citenamefont {de~Léséleuc}, \citenamefont {Macrì}, \citenamefont {Lahaye},\ and\ \citenamefont {Browaeys}}]{Labuhn2016}%
  \BibitemOpen
  \bibfield  {author} {\bibinfo {author} {\bibfnamefont {H.}~\bibnamefont {Labuhn}}, \bibinfo {author} {\bibfnamefont {D.}~\bibnamefont {Barredo}}, \bibinfo {author} {\bibfnamefont {S.}~\bibnamefont {Ravets}}, \bibinfo {author} {\bibfnamefont {S.}~\bibnamefont {de~Léséleuc}}, \bibinfo {author} {\bibfnamefont {T.}~\bibnamefont {Macrì}}, \bibinfo {author} {\bibfnamefont {T.}~\bibnamefont {Lahaye}},\ and\ \bibinfo {author} {\bibfnamefont {A.}~\bibnamefont {Browaeys}},\ }\href {https://doi.org/10.1038/nature18274} {\bibfield  {journal} {\bibinfo  {journal} {Nature}\ }\textbf {\bibinfo {volume} {534}},\ \bibinfo {pages} {667} (\bibinfo {year} {2016})}\BibitemShut {NoStop}%
\bibitem [{\citenamefont {Campbell}\ \emph {et~al.}(2017)\citenamefont {Campbell}, \citenamefont {Hutson}, \citenamefont {Marti}, \citenamefont {Goban}, \citenamefont {Oppong}, \citenamefont {McNally}, \citenamefont {Sonderhouse}, \citenamefont {Robinson}, \citenamefont {Zhang}, \citenamefont {Bloom},\ and\ \citenamefont {Ye}}]{CampbellYe2017}%
  \BibitemOpen
  \bibfield  {author} {\bibinfo {author} {\bibfnamefont {S.~L.}\ \bibnamefont {Campbell}}, \bibinfo {author} {\bibfnamefont {R.~B.}\ \bibnamefont {Hutson}}, \bibinfo {author} {\bibfnamefont {G.~E.}\ \bibnamefont {Marti}}, \bibinfo {author} {\bibfnamefont {A.}~\bibnamefont {Goban}}, \bibinfo {author} {\bibfnamefont {N.~D.}\ \bibnamefont {Oppong}}, \bibinfo {author} {\bibfnamefont {R.~L.}\ \bibnamefont {McNally}}, \bibinfo {author} {\bibfnamefont {L.}~\bibnamefont {Sonderhouse}}, \bibinfo {author} {\bibfnamefont {J.~M.}\ \bibnamefont {Robinson}}, \bibinfo {author} {\bibfnamefont {W.}~\bibnamefont {Zhang}}, \bibinfo {author} {\bibfnamefont {B.~J.}\ \bibnamefont {Bloom}},\ and\ \bibinfo {author} {\bibfnamefont {J.}~\bibnamefont {Ye}},\ }\href {https://doi.org/10.1126/science.aam5538} {\bibfield  {journal} {\bibinfo  {journal} {Science}\ }\textbf {\bibinfo {volume} {358}},\ \bibinfo {pages} {90} (\bibinfo {year} {2017})}\BibitemShut {NoStop}%
\bibitem [{\citenamefont {Hutson}\ \emph {et~al.}(2019)\citenamefont {Hutson}, \citenamefont {Goban}, \citenamefont {Marti}, \citenamefont {Sonderhouse}, \citenamefont {Sanner},\ and\ \citenamefont {Ye}}]{HutsonYe2019}%
  \BibitemOpen
  \bibfield  {author} {\bibinfo {author} {\bibfnamefont {R.~B.}\ \bibnamefont {Hutson}}, \bibinfo {author} {\bibfnamefont {A.}~\bibnamefont {Goban}}, \bibinfo {author} {\bibfnamefont {G.~E.}\ \bibnamefont {Marti}}, \bibinfo {author} {\bibfnamefont {L.}~\bibnamefont {Sonderhouse}}, \bibinfo {author} {\bibfnamefont {C.}~\bibnamefont {Sanner}},\ and\ \bibinfo {author} {\bibfnamefont {J.}~\bibnamefont {Ye}},\ }\href {https://doi.org/10.1103/PhysRevLett.123.123401} {\bibfield  {journal} {\bibinfo  {journal} {Phys. Rev. Lett.}\ }\textbf {\bibinfo {volume} {123}},\ \bibinfo {pages} {123401} (\bibinfo {year} {2019})}\BibitemShut {NoStop}%
\bibitem [{\citenamefont {Beloy}\ \emph {et~al.}(2021)\citenamefont {Beloy}, \citenamefont {Bodine}, \citenamefont {Bothwell}, \citenamefont {Brewer}, \citenamefont {Bromley}, \citenamefont {Chen}, \citenamefont {Desch{\^e}nes}, \citenamefont {Diddams}, \citenamefont {Fasano}, \citenamefont {Fortier}, \citenamefont {Hassan}, \citenamefont {Hume}, \citenamefont {Kedar}, \citenamefont {Kennedy}, \citenamefont {Khader}, \citenamefont {Koepke}, \citenamefont {Leibrandt}, \citenamefont {Leopardi}, \citenamefont {Ludlow}, \citenamefont {McGrew}, \citenamefont {Milner}, \citenamefont {Newbury}, \citenamefont {Nicolodi}, \citenamefont {Oelker}, \citenamefont {Parker}, \citenamefont {Robinson}, \citenamefont {Romisch}, \citenamefont {Sch{\"a}ffer}, \citenamefont {Sherman}, \citenamefont {Sinclair}, \citenamefont {Sonderhouse}, \citenamefont {Swann}, \citenamefont {Yao}, \citenamefont {Ye},\ and\ \citenamefont {Zhang}}]{BACON_collab_2021}%
  \BibitemOpen
  \bibfield  {author} {\bibinfo {author} {\bibfnamefont {K.}~\bibnamefont {Beloy}}, \bibinfo {author} {\bibfnamefont {M.~I.}\ \bibnamefont {Bodine}}, \bibinfo {author} {\bibfnamefont {T.}~\bibnamefont {Bothwell}}, \bibinfo {author} {\bibfnamefont {S.~M.}\ \bibnamefont {Brewer}}, \bibinfo {author} {\bibfnamefont {S.~L.}\ \bibnamefont {Bromley}}, \bibinfo {author} {\bibfnamefont {J.-S.}\ \bibnamefont {Chen}}, \bibinfo {author} {\bibfnamefont {J.-D.}\ \bibnamefont {Desch{\^e}nes}}, \bibinfo {author} {\bibfnamefont {S.~A.}\ \bibnamefont {Diddams}}, \bibinfo {author} {\bibfnamefont {R.~J.}\ \bibnamefont {Fasano}}, \bibinfo {author} {\bibfnamefont {T.~M.}\ \bibnamefont {Fortier}}, \bibinfo {author} {\bibfnamefont {Y.~S.}\ \bibnamefont {Hassan}}, \bibinfo {author} {\bibfnamefont {D.~B.}\ \bibnamefont {Hume}}, \bibinfo {author} {\bibfnamefont {D.}~\bibnamefont {Kedar}}, \bibinfo {author} {\bibfnamefont {C.~J.}\ \bibnamefont {Kennedy}}, \bibinfo {author} {\bibfnamefont {I.}~\bibnamefont {Khader}}, \bibinfo {author} {\bibfnamefont {A.}~\bibnamefont {Koepke}}, \bibinfo {author} {\bibfnamefont {D.~R.}\ \bibnamefont {Leibrandt}}, \bibinfo {author} {\bibfnamefont {H.}~\bibnamefont {Leopardi}}, \bibinfo {author} {\bibfnamefont {A.~D.}\ \bibnamefont {Ludlow}}, \bibinfo {author} {\bibfnamefont {W.~F.}\ \bibnamefont {McGrew}}, \bibinfo {author} {\bibfnamefont {W.~R.}\ \bibnamefont {Milner}}, \bibinfo {author} {\bibfnamefont {N.~R.}\ \bibnamefont {Newbury}}, \bibinfo {author} {\bibfnamefont {D.}~\bibnamefont {Nicolodi}}, \bibinfo {author} {\bibfnamefont {E.}~\bibnamefont {Oelker}}, \bibinfo {author} {\bibfnamefont {T.~E.}\ \bibnamefont {Parker}}, \bibinfo {author} {\bibfnamefont {J.~M.}\ \bibnamefont {Robinson}}, \bibinfo {author} {\bibfnamefont {S.}~\bibnamefont {Romisch}}, \bibinfo {author} {\bibfnamefont {S.~A.}\ \bibnamefont {Sch{\"a}ffer}}, \bibinfo {author} {\bibfnamefont {J.~A.}\ \bibnamefont {Sherman}}, \bibinfo {author} {\bibfnamefont {L.~C.}\ \bibnamefont {Sinclair}}, \bibinfo {author} {\bibfnamefont {L.}~\bibnamefont {Sonderhouse}}, \bibinfo {author} {\bibfnamefont {W.~C.}\ \bibnamefont {Swann}}, \bibinfo {author} {\bibfnamefont {J.}~\bibnamefont {Yao}}, \bibinfo {author} {\bibfnamefont {J.}~\bibnamefont {Ye}},\ and\ \bibinfo {author} {\bibfnamefont {X.}~\bibnamefont {Zhang}},\ }\href {https://doi.org/10.1038/s41586-021-03253-4} {\bibfield  {journal} {\bibinfo  {journal} {Nature}\ }\textbf {\bibinfo {volume} {591}},\ \bibinfo {pages} {564} (\bibinfo {year} {2021})}\BibitemShut {NoStop}%
\bibitem [{\citenamefont {Bornet}\ \emph {et~al.}(2023{\natexlab{b}})\citenamefont {Bornet}, \citenamefont {Emperauger}, \citenamefont {Chen}, \citenamefont {Ye}, \citenamefont {Block}, \citenamefont {Bintz}, \citenamefont {Boyd}, \citenamefont {Barredo}, \citenamefont {Comparin}, \citenamefont {Mezzacapo} \emph {et~al.}}]{bornet2023}%
  \BibitemOpen
  \bibfield  {author} {\bibinfo {author} {\bibfnamefont {G.}~\bibnamefont {Bornet}}, \bibinfo {author} {\bibfnamefont {G.}~\bibnamefont {Emperauger}}, \bibinfo {author} {\bibfnamefont {C.}~\bibnamefont {Chen}}, \bibinfo {author} {\bibfnamefont {B.}~\bibnamefont {Ye}}, \bibinfo {author} {\bibfnamefont {M.}~\bibnamefont {Block}}, \bibinfo {author} {\bibfnamefont {M.}~\bibnamefont {Bintz}}, \bibinfo {author} {\bibfnamefont {J.~A.}\ \bibnamefont {Boyd}}, \bibinfo {author} {\bibfnamefont {D.}~\bibnamefont {Barredo}}, \bibinfo {author} {\bibfnamefont {T.}~\bibnamefont {Comparin}}, \bibinfo {author} {\bibfnamefont {F.}~\bibnamefont {Mezzacapo}}, \emph {et~al.},\ }\href@noop {} {\bibfield  {journal} {\bibinfo  {journal} {Nature}\ }\textbf {\bibinfo {volume} {621}},\ \bibinfo {pages} {728} (\bibinfo {year} {2023}{\natexlab{b}})}\BibitemShut {NoStop}%
\bibitem [{\citenamefont {Hashizume}\ \emph {et~al.}(2021)\citenamefont {Hashizume}, \citenamefont {Bentsen}, \citenamefont {Weber},\ and\ \citenamefont {Daley}}]{Tomohiro_deter_2021}%
  \BibitemOpen
  \bibfield  {author} {\bibinfo {author} {\bibfnamefont {T.}~\bibnamefont {Hashizume}}, \bibinfo {author} {\bibfnamefont {G.~S.}\ \bibnamefont {Bentsen}}, \bibinfo {author} {\bibfnamefont {S.}~\bibnamefont {Weber}},\ and\ \bibinfo {author} {\bibfnamefont {A.~J.}\ \bibnamefont {Daley}},\ }\href {https://doi.org/10.1103/PhysRevLett.126.200603} {\bibfield  {journal} {\bibinfo  {journal} {Phys. Rev. Lett.}\ }\textbf {\bibinfo {volume} {126}},\ \bibinfo {pages} {200603} (\bibinfo {year} {2021})}\BibitemShut {NoStop}%
\bibitem [{\citenamefont {Bentsen}\ \emph {et~al.}(2019)\citenamefont {Bentsen}, \citenamefont {Hashizume}, \citenamefont {Buyskikh}, \citenamefont {Davis}, \citenamefont {Daley}, \citenamefont {Gubser},\ and\ \citenamefont {Schleier-Smith}}]{Greg_2019_tree}%
  \BibitemOpen
  \bibfield  {author} {\bibinfo {author} {\bibfnamefont {G.}~\bibnamefont {Bentsen}}, \bibinfo {author} {\bibfnamefont {T.}~\bibnamefont {Hashizume}}, \bibinfo {author} {\bibfnamefont {A.~S.}\ \bibnamefont {Buyskikh}}, \bibinfo {author} {\bibfnamefont {E.~J.}\ \bibnamefont {Davis}}, \bibinfo {author} {\bibfnamefont {A.~J.}\ \bibnamefont {Daley}}, \bibinfo {author} {\bibfnamefont {S.~S.}\ \bibnamefont {Gubser}},\ and\ \bibinfo {author} {\bibfnamefont {M.}~\bibnamefont {Schleier-Smith}},\ }\href {https://doi.org/10.1103/PhysRevLett.123.130601} {\bibfield  {journal} {\bibinfo  {journal} {Phys. Rev. Lett.}\ }\textbf {\bibinfo {volume} {123}},\ \bibinfo {pages} {130601} (\bibinfo {year} {2019})}\BibitemShut {NoStop}%
\bibitem [{\citenamefont {Gubser}\ \emph {et~al.}(2018)\citenamefont {Gubser}, \citenamefont {Jepsen}, \citenamefont {Ji},\ and\ \citenamefont {Trundy}}]{Gubser_2018_continuum}%
  \BibitemOpen
  \bibfield  {author} {\bibinfo {author} {\bibfnamefont {S.~S.}\ \bibnamefont {Gubser}}, \bibinfo {author} {\bibfnamefont {C.}~\bibnamefont {Jepsen}}, \bibinfo {author} {\bibfnamefont {Z.}~\bibnamefont {Ji}},\ and\ \bibinfo {author} {\bibfnamefont {B.}~\bibnamefont {Trundy}},\ }\href {https://doi.org/10.1103/PhysRevD.98.045009} {\bibfield  {journal} {\bibinfo  {journal} {Phys. Rev. D}\ }\textbf {\bibinfo {volume} {98}},\ \bibinfo {pages} {045009} (\bibinfo {year} {2018})}\BibitemShut {NoStop}%
\bibitem [{\citenamefont {Gubser}\ \emph {et~al.}(2019)\citenamefont {Gubser}, \citenamefont {Jepsen}, \citenamefont {Ji},\ and\ \citenamefont {Trundy}}]{Gubser_2019_mixed}%
  \BibitemOpen
  \bibfield  {author} {\bibinfo {author} {\bibfnamefont {S.~S.}\ \bibnamefont {Gubser}}, \bibinfo {author} {\bibfnamefont {C.}~\bibnamefont {Jepsen}}, \bibinfo {author} {\bibfnamefont {Z.}~\bibnamefont {Ji}},\ and\ \bibinfo {author} {\bibfnamefont {B.}~\bibnamefont {Trundy}},\ }\href {https://doi.org/10.1007/jhep12(2019)136} {\bibinfo  {journal} {J. High Energ. Phys.}\ ,\ \bibinfo {pages} {1}}\BibitemShut {NoStop}%
\bibitem [{\citenamefont {Kuriyattil}\ \emph {et~al.}(2023)\citenamefont {Kuriyattil}, \citenamefont {Hashizume}, \citenamefont {Bentsen},\ and\ \citenamefont {Daley}}]{Kuriyattil_2023}%
  \BibitemOpen
\bibfield  {journal} {  }\bibfield  {author} {\bibinfo {author} {\bibfnamefont {S.}~\bibnamefont {Kuriyattil}}, \bibinfo {author} {\bibfnamefont {T.}~\bibnamefont {Hashizume}}, \bibinfo {author} {\bibfnamefont {G.}~\bibnamefont {Bentsen}},\ and\ \bibinfo {author} {\bibfnamefont {A.~J.}\ \bibnamefont {Daley}},\ }\href {https://doi.org/10.1103/PRXQuantum.4.030325} {\bibfield  {journal} {\bibinfo  {journal} {PRX Quantum}\ }\textbf {\bibinfo {volume} {4}},\ \bibinfo {pages} {030325} (\bibinfo {year} {2023})}\BibitemShut {NoStop}%
\bibitem [{\citenamefont {Hashizume}\ \emph {et~al.}(2022)\citenamefont {Hashizume}, \citenamefont {Kuriyattil}, \citenamefont {Daley},\ and\ \citenamefont {Bentsen}}]{Hashizume_2022}%
  \BibitemOpen
  \bibfield  {author} {\bibinfo {author} {\bibfnamefont {T.}~\bibnamefont {Hashizume}}, \bibinfo {author} {\bibfnamefont {S.}~\bibnamefont {Kuriyattil}}, \bibinfo {author} {\bibfnamefont {A.~J.}\ \bibnamefont {Daley}},\ and\ \bibinfo {author} {\bibfnamefont {G.}~\bibnamefont {Bentsen}},\ }\href {https://doi.org/10.3390/sym14040666} {\bibfield  {journal} {\bibinfo  {journal} {Symmetry}\ }\textbf {\bibinfo {volume} {14}},\ \bibinfo {pages} {666} (\bibinfo {year} {2022})}\BibitemShut {NoStop}%
\bibitem [{\citenamefont {Periwal}\ \emph {et~al.}(2021)\citenamefont {Periwal}, \citenamefont {Cooper}, \citenamefont {Kunkel}, \citenamefont {Wienand}, \citenamefont {Davis},\ and\ \citenamefont {Schleier-Smith}}]{periwal_programmable_2021}%
  \BibitemOpen
  \bibfield  {author} {\bibinfo {author} {\bibfnamefont {A.}~\bibnamefont {Periwal}}, \bibinfo {author} {\bibfnamefont {E.~S.}\ \bibnamefont {Cooper}}, \bibinfo {author} {\bibfnamefont {P.}~\bibnamefont {Kunkel}}, \bibinfo {author} {\bibfnamefont {J.~F.}\ \bibnamefont {Wienand}}, \bibinfo {author} {\bibfnamefont {E.~J.}\ \bibnamefont {Davis}},\ and\ \bibinfo {author} {\bibfnamefont {M.}~\bibnamefont {Schleier-Smith}},\ }\href {https://doi.org/10.1038/s41586-021-04156-0} {\bibfield  {journal} {\bibinfo  {journal} {Nature}\ }\textbf {\bibinfo {volume} {600}},\ \bibinfo {pages} {630} (\bibinfo {year} {2021})}\BibitemShut {NoStop}%
\bibitem [{\citenamefont {Barredo}\ \emph {et~al.}(2016)\citenamefont {Barredo}, \citenamefont {de~Léséleuc}, \citenamefont {Lienhard}, \citenamefont {Lahaye},\ and\ \citenamefont {Browaeys}}]{Barredo_2016}%
  \BibitemOpen
  \bibfield  {author} {\bibinfo {author} {\bibfnamefont {D.}~\bibnamefont {Barredo}}, \bibinfo {author} {\bibfnamefont {S.}~\bibnamefont {de~Léséleuc}}, \bibinfo {author} {\bibfnamefont {V.}~\bibnamefont {Lienhard}}, \bibinfo {author} {\bibfnamefont {T.}~\bibnamefont {Lahaye}},\ and\ \bibinfo {author} {\bibfnamefont {A.}~\bibnamefont {Browaeys}},\ }\href {https://doi.org/10.1126/science.aah3778} {\bibfield  {journal} {\bibinfo  {journal} {Science}\ }\textbf {\bibinfo {volume} {354}},\ \bibinfo {pages} {1021–1023} (\bibinfo {year} {2016})}\BibitemShut {NoStop}%
\bibitem [{\citenamefont {Kim}\ \emph {et~al.}(2016)\citenamefont {Kim}, \citenamefont {Lee}, \citenamefont {Lee}, \citenamefont {Jo}, \citenamefont {Song},\ and\ \citenamefont {Ahn}}]{Kim_2016}%
  \BibitemOpen
  \bibfield  {author} {\bibinfo {author} {\bibfnamefont {H.}~\bibnamefont {Kim}}, \bibinfo {author} {\bibfnamefont {W.}~\bibnamefont {Lee}}, \bibinfo {author} {\bibfnamefont {H.-g.}\ \bibnamefont {Lee}}, \bibinfo {author} {\bibfnamefont {H.}~\bibnamefont {Jo}}, \bibinfo {author} {\bibfnamefont {Y.}~\bibnamefont {Song}},\ and\ \bibinfo {author} {\bibfnamefont {J.}~\bibnamefont {Ahn}},\ }\href {https://www.nature.com/articles/ncomms13317} {\bibfield  {journal} {\bibinfo  {journal} {Nature Communications}\ }\textbf {\bibinfo {volume} {7}} (\bibinfo {year} {2016})}\BibitemShut {NoStop}%
\bibitem [{\citenamefont {Endres}\ \emph {et~al.}(2016)\citenamefont {Endres}, \citenamefont {Bernien}, \citenamefont {Keesling}, \citenamefont {Levine}, \citenamefont {Anschuetz}, \citenamefont {Krajenbrink}, \citenamefont {Senko}, \citenamefont {Vuletic}, \citenamefont {Greiner},\ and\ \citenamefont {Lukin}}]{Manuel_atom_assembly_2016}%
  \BibitemOpen
  \bibfield  {author} {\bibinfo {author} {\bibfnamefont {M.}~\bibnamefont {Endres}}, \bibinfo {author} {\bibfnamefont {H.}~\bibnamefont {Bernien}}, \bibinfo {author} {\bibfnamefont {A.}~\bibnamefont {Keesling}}, \bibinfo {author} {\bibfnamefont {H.}~\bibnamefont {Levine}}, \bibinfo {author} {\bibfnamefont {E.~R.}\ \bibnamefont {Anschuetz}}, \bibinfo {author} {\bibfnamefont {A.}~\bibnamefont {Krajenbrink}}, \bibinfo {author} {\bibfnamefont {C.}~\bibnamefont {Senko}}, \bibinfo {author} {\bibfnamefont {V.}~\bibnamefont {Vuletic}}, \bibinfo {author} {\bibfnamefont {M.}~\bibnamefont {Greiner}},\ and\ \bibinfo {author} {\bibfnamefont {M.~D.}\ \bibnamefont {Lukin}},\ }\href {https://doi.org/10.1126/science.aah3752} {\bibfield  {journal} {\bibinfo  {journal} {Science}\ }\textbf {\bibinfo {volume} {354}},\ \bibinfo {pages} {1024} (\bibinfo {year} {2016})}\BibitemShut {NoStop}%
\bibitem [{\citenamefont {Mu{\~n}oz-Arias}\ \emph {et~al.}(2023)\citenamefont {Mu{\~n}oz-Arias}, \citenamefont {Deutsch},\ and\ \citenamefont {Poggi}}]{munoz2023}%
  \BibitemOpen
  \bibfield  {author} {\bibinfo {author} {\bibfnamefont {M.~H.}\ \bibnamefont {Mu{\~n}oz-Arias}}, \bibinfo {author} {\bibfnamefont {I.~H.}\ \bibnamefont {Deutsch}},\ and\ \bibinfo {author} {\bibfnamefont {P.~M.}\ \bibnamefont {Poggi}},\ }\href {https://doi.org/10.1103/PRXQuantum.4.020314} {\bibfield  {journal} {\bibinfo  {journal} {PRX Quantum}\ }\textbf {\bibinfo {volume} {4}},\ \bibinfo {pages} {020314} (\bibinfo {year} {2023})}\BibitemShut {NoStop}%
\bibitem [{\citenamefont {Britton}\ \emph {et~al.}(2012)\citenamefont {Britton}, \citenamefont {Sawyer}, \citenamefont {Keith}, \citenamefont {Wang}, \citenamefont {Freericks}, \citenamefont {Uys}, \citenamefont {Biercuk},\ and\ \citenamefont {Bollinger}}]{Britton_2012}%
  \BibitemOpen
  \bibfield  {author} {\bibinfo {author} {\bibfnamefont {J.~W.}\ \bibnamefont {Britton}}, \bibinfo {author} {\bibfnamefont {B.~C.}\ \bibnamefont {Sawyer}}, \bibinfo {author} {\bibfnamefont {A.~C.}\ \bibnamefont {Keith}}, \bibinfo {author} {\bibfnamefont {C.-C.~J.}\ \bibnamefont {Wang}}, \bibinfo {author} {\bibfnamefont {J.~K.}\ \bibnamefont {Freericks}}, \bibinfo {author} {\bibfnamefont {H.}~\bibnamefont {Uys}}, \bibinfo {author} {\bibfnamefont {M.~J.}\ \bibnamefont {Biercuk}},\ and\ \bibinfo {author} {\bibfnamefont {J.~J.}\ \bibnamefont {Bollinger}},\ }\href {https://doi.org/10.1038/nature10981} {\bibfield  {journal} {\bibinfo  {journal} {Nature}\ }\textbf {\bibinfo {volume} {484}},\ \bibinfo {pages} {489–492} (\bibinfo {year} {2012})}\BibitemShut {NoStop}%
\bibitem [{\citenamefont {Bohnet}\ \emph {et~al.}(2016)\citenamefont {Bohnet}, \citenamefont {Sawyer}, \citenamefont {Britton}, \citenamefont {Wall}, \citenamefont {Rey}, \citenamefont {Foss-Feig},\ and\ \citenamefont {Bollinger}}]{Justin_qsd_2016}%
  \BibitemOpen
  \bibfield  {author} {\bibinfo {author} {\bibfnamefont {J.~G.}\ \bibnamefont {Bohnet}}, \bibinfo {author} {\bibfnamefont {B.~C.}\ \bibnamefont {Sawyer}}, \bibinfo {author} {\bibfnamefont {J.~W.}\ \bibnamefont {Britton}}, \bibinfo {author} {\bibfnamefont {M.~L.}\ \bibnamefont {Wall}}, \bibinfo {author} {\bibfnamefont {A.~M.}\ \bibnamefont {Rey}}, \bibinfo {author} {\bibfnamefont {M.}~\bibnamefont {Foss-Feig}},\ and\ \bibinfo {author} {\bibfnamefont {J.~J.}\ \bibnamefont {Bollinger}},\ }\href {https://doi.org/10.1126/science.aad9958} {\bibfield  {journal} {\bibinfo  {journal} {Science}\ }\textbf {\bibinfo {volume} {352}},\ \bibinfo {pages} {1297} (\bibinfo {year} {2016})}\BibitemShut {NoStop}%
\bibitem [{\citenamefont {Davis}\ \emph {et~al.}(2019)\citenamefont {Davis}, \citenamefont {Bentsen}, \citenamefont {Homeier}, \citenamefont {Li},\ and\ \citenamefont {Schleier-Smith}}]{Emily_Photon_mdeiated_2019}%
  \BibitemOpen
  \bibfield  {author} {\bibinfo {author} {\bibfnamefont {E.~J.}\ \bibnamefont {Davis}}, \bibinfo {author} {\bibfnamefont {G.}~\bibnamefont {Bentsen}}, \bibinfo {author} {\bibfnamefont {L.}~\bibnamefont {Homeier}}, \bibinfo {author} {\bibfnamefont {T.}~\bibnamefont {Li}},\ and\ \bibinfo {author} {\bibfnamefont {M.~H.}\ \bibnamefont {Schleier-Smith}},\ }\href {https://doi.org/10.1103/PhysRevLett.122.010405} {\bibfield  {journal} {\bibinfo  {journal} {Phys. Rev. Lett.}\ }\textbf {\bibinfo {volume} {122}},\ \bibinfo {pages} {010405} (\bibinfo {year} {2019})}\BibitemShut {NoStop}%
\bibitem [{\citenamefont {Ritsch}\ \emph {et~al.}(2013)\citenamefont {Ritsch}, \citenamefont {Domokos}, \citenamefont {Brennecke},\ and\ \citenamefont {Esslinger}}]{Ritsch_cold_atoms_2013}%
  \BibitemOpen
  \bibfield  {author} {\bibinfo {author} {\bibfnamefont {H.}~\bibnamefont {Ritsch}}, \bibinfo {author} {\bibfnamefont {P.}~\bibnamefont {Domokos}}, \bibinfo {author} {\bibfnamefont {F.}~\bibnamefont {Brennecke}},\ and\ \bibinfo {author} {\bibfnamefont {T.}~\bibnamefont {Esslinger}},\ }\href {https://doi.org/10.1103/RevModPhys.85.553} {\bibfield  {journal} {\bibinfo  {journal} {Rev. Mod. Phys.}\ }\textbf {\bibinfo {volume} {85}},\ \bibinfo {pages} {553} (\bibinfo {year} {2013})}\BibitemShut {NoStop}%
\bibitem [{\citenamefont {Baumann}\ \emph {et~al.}(2010)\citenamefont {Baumann}, \citenamefont {Guerlin}, \citenamefont {Brennecke},\ and\ \citenamefont {Esslinger}}]{Baumann_2010}%
  \BibitemOpen
  \bibfield  {author} {\bibinfo {author} {\bibfnamefont {K.}~\bibnamefont {Baumann}}, \bibinfo {author} {\bibfnamefont {C.}~\bibnamefont {Guerlin}}, \bibinfo {author} {\bibfnamefont {F.}~\bibnamefont {Brennecke}},\ and\ \bibinfo {author} {\bibfnamefont {T.}~\bibnamefont {Esslinger}},\ }\href {https://doi.org/10.1038/nature09009} {\bibfield  {journal} {\bibinfo  {journal} {Nature}\ }\textbf {\bibinfo {volume} {464}},\ \bibinfo {pages} {1301–1306} (\bibinfo {year} {2010})}\BibitemShut {NoStop}%
\bibitem [{\citenamefont {Perlin}\ \emph {et~al.}(2020)\citenamefont {Perlin}, \citenamefont {Qu},\ and\ \citenamefont {Rey}}]{perlin2020}%
  \BibitemOpen
  \bibfield  {author} {\bibinfo {author} {\bibfnamefont {M.~A.}\ \bibnamefont {Perlin}}, \bibinfo {author} {\bibfnamefont {C.}~\bibnamefont {Qu}},\ and\ \bibinfo {author} {\bibfnamefont {A.~M.}\ \bibnamefont {Rey}},\ }\href {https://doi.org/10.1103/PhysRevLett.125.223401} {\bibfield  {journal} {\bibinfo  {journal} {Phys. Rev. Lett.}\ }\textbf {\bibinfo {volume} {125}},\ \bibinfo {pages} {223401} (\bibinfo {year} {2020})}\BibitemShut {NoStop}%
\bibitem [{\citenamefont {Block}\ \emph {et~al.}(2024)\citenamefont {Block}, \citenamefont {Ye}, \citenamefont {Roberts}, \citenamefont {Chern}, \citenamefont {Wu}, \citenamefont {Wang}, \citenamefont {Pollet}, \citenamefont {Davis}, \citenamefont {Halperin},\ and\ \citenamefont {Yao}}]{block2023universal}%
  \BibitemOpen
  \bibfield  {author} {\bibinfo {author} {\bibfnamefont {M.}~\bibnamefont {Block}}, \bibinfo {author} {\bibfnamefont {B.}~\bibnamefont {Ye}}, \bibinfo {author} {\bibfnamefont {B.}~\bibnamefont {Roberts}}, \bibinfo {author} {\bibfnamefont {S.}~\bibnamefont {Chern}}, \bibinfo {author} {\bibfnamefont {W.}~\bibnamefont {Wu}}, \bibinfo {author} {\bibfnamefont {Z.}~\bibnamefont {Wang}}, \bibinfo {author} {\bibfnamefont {L.}~\bibnamefont {Pollet}}, \bibinfo {author} {\bibfnamefont {E.~J.}\ \bibnamefont {Davis}}, \bibinfo {author} {\bibfnamefont {B.~I.}\ \bibnamefont {Halperin}},\ and\ \bibinfo {author} {\bibfnamefont {N.~Y.}\ \bibnamefont {Yao}},\ }\href {https://doi.org/10.1038/s41567-024-02562-5} {\bibfield  {journal} {\bibinfo  {journal} {Nature Physics}\ } (\bibinfo {year} {2024})}\BibitemShut {NoStop}%
\bibitem [{\citenamefont {Pezz\`e}\ \emph {et~al.}(2018)\citenamefont {Pezz\`e}, \citenamefont {Smerzi}, \citenamefont {Oberthaler}, \citenamefont {Schmied},\ and\ \citenamefont {Treutlein}}]{PezzeTreutlein2018}%
  \BibitemOpen
  \bibfield  {author} {\bibinfo {author} {\bibfnamefont {L.}~\bibnamefont {Pezz\`e}}, \bibinfo {author} {\bibfnamefont {A.}~\bibnamefont {Smerzi}}, \bibinfo {author} {\bibfnamefont {M.~K.}\ \bibnamefont {Oberthaler}}, \bibinfo {author} {\bibfnamefont {R.}~\bibnamefont {Schmied}},\ and\ \bibinfo {author} {\bibfnamefont {P.}~\bibnamefont {Treutlein}},\ }\href {https://doi.org/10.1103/RevModPhys.90.035005} {\bibfield  {journal} {\bibinfo  {journal} {Rev. Mod. Phys.}\ }\textbf {\bibinfo {volume} {90}},\ \bibinfo {pages} {035005} (\bibinfo {year} {2018})}\BibitemShut {NoStop}%
\bibitem [{\citenamefont {Pezz{\`e}}\ and\ \citenamefont {Smerzi}(2014)}]{PezzeSmerzi2014}%
  \BibitemOpen
  \bibfield  {author} {\bibinfo {author} {\bibfnamefont {L.}~\bibnamefont {Pezz{\`e}}}\ and\ \bibinfo {author} {\bibfnamefont {A.}~\bibnamefont {Smerzi}},\ }\href {https://api.semanticscholar.org/CorpusID:119223464} {\bibfield  {journal} {\bibinfo  {journal} {arXiv: Quantum Physics}\ } (\bibinfo {year} {2014})}\BibitemShut {NoStop}%
\bibitem [{\citenamefont {Braunstein}\ and\ \citenamefont {Caves}(1994)}]{BraunsteinCaves1994}%
  \BibitemOpen
  \bibfield  {author} {\bibinfo {author} {\bibfnamefont {S.~L.}\ \bibnamefont {Braunstein}}\ and\ \bibinfo {author} {\bibfnamefont {C.~M.}\ \bibnamefont {Caves}},\ }\href {https://doi.org/10.1103/PhysRevLett.72.3439} {\bibfield  {journal} {\bibinfo  {journal} {Phys. Rev. Lett.}\ }\textbf {\bibinfo {volume} {72}},\ \bibinfo {pages} {3439} (\bibinfo {year} {1994})}\BibitemShut {NoStop}%
\bibitem [{\citenamefont {Paris}(2009)}]{Paris2009}%
  \BibitemOpen
  \bibfield  {author} {\bibinfo {author} {\bibfnamefont {M.~G.~A.}\ \bibnamefont {Paris}},\ }\href {https://doi.org/10.1142/S0219749909004839} {\bibfield  {journal} {\bibinfo  {journal} {International Journal of Quantum Information}\ }\textbf {\bibinfo {volume} {07}},\ \bibinfo {pages} {125} (\bibinfo {year} {2009})}\BibitemShut {NoStop}%
\bibitem [{\citenamefont {Tóth}\ and\ \citenamefont {Apellaniz}(2014)}]{TothApellaniz2014}%
  \BibitemOpen
  \bibfield  {author} {\bibinfo {author} {\bibfnamefont {G.}~\bibnamefont {Tóth}}\ and\ \bibinfo {author} {\bibfnamefont {I.}~\bibnamefont {Apellaniz}},\ }\href {https://doi.org/10.1088/1751-8113/47/42/424006} {\bibfield  {journal} {\bibinfo  {journal} {Journal of Physics A: Mathematical and Theoretical}\ }\textbf {\bibinfo {volume} {47}},\ \bibinfo {pages} {424006} (\bibinfo {year} {2014})}\BibitemShut {NoStop}%
\bibitem [{\citenamefont {Helstrom}(1969)}]{Helstrom1969}%
  \BibitemOpen
  \bibfield  {author} {\bibinfo {author} {\bibfnamefont {C.~W.}\ \bibnamefont {Helstrom}},\ }\href {https://doi.org/10.1007/BF01007479} {\bibfield  {journal} {\bibinfo  {journal} {Journal of Statistical Physics}\ }\textbf {\bibinfo {volume} {1}},\ \bibinfo {pages} {231} (\bibinfo {year} {1969})}\BibitemShut {NoStop}%
\bibitem [{Sup()}]{SupMat}%
  \BibitemOpen
  \href@noop {} {\emph {\bibinfo {title} {{See Supplemental Material}}}}\BibitemShut {NoStop}%
\bibitem [{\citenamefont {White}\ and\ \citenamefont {Feiguin}(2004)}]{WhiteFeiguin2004}%
  \BibitemOpen
  \bibfield  {author} {\bibinfo {author} {\bibfnamefont {S.~R.}\ \bibnamefont {White}}\ and\ \bibinfo {author} {\bibfnamefont {A.~E.}\ \bibnamefont {Feiguin}},\ }\href {https://doi.org/10.1103/PhysRevLett.93.076401} {\bibfield  {journal} {\bibinfo  {journal} {Phys. Rev. Lett.}\ }\textbf {\bibinfo {volume} {93}},\ \bibinfo {pages} {076401} (\bibinfo {year} {2004})}\BibitemShut {NoStop}%
\bibitem [{\citenamefont {Daley}\ \emph {et~al.}(2004)\citenamefont {Daley}, \citenamefont {Kollath}, \citenamefont {Schollwöck},\ and\ \citenamefont {Vidal}}]{DaleySchollwoeck2004}%
  \BibitemOpen
  \bibfield  {author} {\bibinfo {author} {\bibfnamefont {A.~J.}\ \bibnamefont {Daley}}, \bibinfo {author} {\bibfnamefont {C.}~\bibnamefont {Kollath}}, \bibinfo {author} {\bibfnamefont {U.}~\bibnamefont {Schollwöck}},\ and\ \bibinfo {author} {\bibfnamefont {G.}~\bibnamefont {Vidal}},\ }\href {https://doi.org/10.1088/1742-5468/2004/04/p04005} {\bibfield  {journal} {\bibinfo  {journal} {Journal of Statistical Mechanics: Theory and Experiment}\ }\textbf {\bibinfo {volume} {2004}},\ \bibinfo {pages} {P04005} (\bibinfo {year} {2004})}\BibitemShut {NoStop}%
\bibitem [{\citenamefont {Schollwöck}(2011)}]{Schollwoeck2011}%
  \BibitemOpen
  \bibfield  {author} {\bibinfo {author} {\bibfnamefont {U.}~\bibnamefont {Schollwöck}},\ }\href {https://doi.org/10.1016/j.aop.2010.09.012} {\bibfield  {journal} {\bibinfo  {journal} {Annals of Physics}\ }\textbf {\bibinfo {volume} {326}},\ \bibinfo {pages} {96–192} (\bibinfo {year} {2011})}\BibitemShut {NoStop}%
\bibitem [{\citenamefont {Haegeman}\ \emph {et~al.}(2016)\citenamefont {Haegeman}, \citenamefont {Lubich}, \citenamefont {Oseledets}, \citenamefont {Vandereycken},\ and\ \citenamefont {Verstraete}}]{HaegemanVerstraete2016}%
  \BibitemOpen
  \bibfield  {author} {\bibinfo {author} {\bibfnamefont {J.}~\bibnamefont {Haegeman}}, \bibinfo {author} {\bibfnamefont {C.}~\bibnamefont {Lubich}}, \bibinfo {author} {\bibfnamefont {I.}~\bibnamefont {Oseledets}}, \bibinfo {author} {\bibfnamefont {B.}~\bibnamefont {Vandereycken}},\ and\ \bibinfo {author} {\bibfnamefont {F.}~\bibnamefont {Verstraete}},\ }\href {https://doi.org/10.1103/PhysRevB.94.165116} {\bibfield  {journal} {\bibinfo  {journal} {Phys. Rev. B}\ }\textbf {\bibinfo {volume} {94}},\ \bibinfo {pages} {165116} (\bibinfo {year} {2016})}\BibitemShut {NoStop}%
\bibitem [{\citenamefont {Fishman}\ \emph {et~al.}(2022{\natexlab{a}})\citenamefont {Fishman}, \citenamefont {White},\ and\ \citenamefont {Stoudenmire}}]{FishmanStoudenmire2022}%
  \BibitemOpen
  \bibfield  {author} {\bibinfo {author} {\bibfnamefont {M.}~\bibnamefont {Fishman}}, \bibinfo {author} {\bibfnamefont {S.~R.}\ \bibnamefont {White}},\ and\ \bibinfo {author} {\bibfnamefont {E.~M.}\ \bibnamefont {Stoudenmire}},\ }\href {https://doi.org/10.21468/SciPostPhysCodeb.4} {\bibfield  {journal} {\bibinfo  {journal} {SciPost Phys. Codebases}\ ,\ \bibinfo {pages} {4}} (\bibinfo {year} {2022}{\natexlab{a}})}\BibitemShut {NoStop}%
\bibitem [{\citenamefont {Fishman}\ \emph {et~al.}(2022{\natexlab{b}})\citenamefont {Fishman}, \citenamefont {White},\ and\ \citenamefont {Stoudenmire}}]{ITensor2022}%
  \BibitemOpen
  \bibfield  {author} {\bibinfo {author} {\bibfnamefont {M.}~\bibnamefont {Fishman}}, \bibinfo {author} {\bibfnamefont {S.~R.}\ \bibnamefont {White}},\ and\ \bibinfo {author} {\bibfnamefont {E.~M.}\ \bibnamefont {Stoudenmire}},\ }\href {https://doi.org/10.21468/SciPostPhysCodeb.4-r0.3} {\bibfield  {journal} {\bibinfo  {journal} {SciPost Phys. Codebases}\ ,\ \bibinfo {pages} {4}} (\bibinfo {year} {2022}{\natexlab{b}})}\BibitemShut {NoStop}%
\bibitem [{\citenamefont {Pires}(2021)}]{Pires_spinwavetheory}%
  \BibitemOpen
  \bibfield  {author} {\bibinfo {author} {\bibfnamefont {A.~S.~T.}\ \bibnamefont {Pires}},\ }\href {https://doi.org/10.1088/978-0-7503-3879-0} {\emph {\bibinfo {title} {Theoretical Tools for Spin Models in Magnetic Systems}}},\ 2053-2563\ (\bibinfo  {publisher} {IOP Publishing},\ \bibinfo {year} {2021})\BibitemShut {NoStop}%
\bibitem [{\citenamefont {Stancel}\ and\ \citenamefont {Prabhakar}(2009)}]{stancel2009spin}%
  \BibitemOpen
  \bibfield  {author} {\bibinfo {author} {\bibfnamefont {D.~D.}\ \bibnamefont {Stancel}}\ and\ \bibinfo {author} {\bibfnamefont {A.}~\bibnamefont {Prabhakar}},\ }\href@noop {} {\emph {\bibinfo {title} {Spin Waves: Theory and Applications}}}\ (\bibinfo  {publisher} {Springer, Berlin},\ \bibinfo {year} {2009})\BibitemShut {NoStop}%
\bibitem [{\citenamefont {Kittel}(1963)}]{kittel1963quantum}%
  \BibitemOpen
  \bibfield  {author} {\bibinfo {author} {\bibfnamefont {C.}~\bibnamefont {Kittel}},\ }\href@noop {} {\emph {\bibinfo {title} {Quantum Theory of Solids}}}\ (\bibinfo  {publisher} {Wiley, New York},\ \bibinfo {year} {1963})\BibitemShut {NoStop}%
\bibitem [{\citenamefont {Aldous}\ and\ \citenamefont {Diaconis}(1986)}]{aldous1986shuffling}%
  \BibitemOpen
  \bibfield  {author} {\bibinfo {author} {\bibfnamefont {D.}~\bibnamefont {Aldous}}\ and\ \bibinfo {author} {\bibfnamefont {P.}~\bibnamefont {Diaconis}},\ }\href {https://www.jstor.org/stable/i315161} {\bibfield  {journal} {\bibinfo  {journal} {The American Mathematical Monthly}\ }\textbf {\bibinfo {volume} {93}},\ \bibinfo {pages} {333} (\bibinfo {year} {1986})}\BibitemShut {NoStop}%
\bibitem [{\citenamefont {Diaconis}\ \emph {et~al.}(1983)\citenamefont {Diaconis}, \citenamefont {Graham},\ and\ \citenamefont {Kantor}}]{diaconis1983mathematics}%
  \BibitemOpen
  \bibfield  {author} {\bibinfo {author} {\bibfnamefont {P.}~\bibnamefont {Diaconis}}, \bibinfo {author} {\bibfnamefont {R.~L.}\ \bibnamefont {Graham}},\ and\ \bibinfo {author} {\bibfnamefont {W.~M.}\ \bibnamefont {Kantor}},\ }\href {https://doi.org/https://doi.org/10.1016/0196-8858(83)90009-X} {\bibfield  {journal} {\bibinfo  {journal} {Advances in Applied Mathematics}\ }\textbf {\bibinfo {volume} {4}},\ \bibinfo {pages} {175} (\bibinfo {year} {1983})}\BibitemShut {NoStop}%
\bibitem [{\citenamefont {Heyl}\ \emph {et~al.}(2019)\citenamefont {Heyl}, \citenamefont {Hauke},\ and\ \citenamefont {Zoller}}]{Heyl_2019}%
  \BibitemOpen
  \bibfield  {author} {\bibinfo {author} {\bibfnamefont {M.}~\bibnamefont {Heyl}}, \bibinfo {author} {\bibfnamefont {P.}~\bibnamefont {Hauke}},\ and\ \bibinfo {author} {\bibfnamefont {P.}~\bibnamefont {Zoller}},\ }\href {http://dx.doi.org/10.1126/sciadv.aau8342} {\bibfield  {journal} {\bibinfo  {journal} {Science Advances}\ }\textbf {\bibinfo {volume} {5}} (\bibinfo {year} {2019})}\BibitemShut {NoStop}%
\bibitem [{\citenamefont {Chinni}\ \emph {et~al.}(2022)\citenamefont {Chinni}, \citenamefont {Mu{\~n}oz-Arias}, \citenamefont {Deutsch},\ and\ \citenamefont {Poggi}}]{chinni2022}%
  \BibitemOpen
  \bibfield  {author} {\bibinfo {author} {\bibfnamefont {K.}~\bibnamefont {Chinni}}, \bibinfo {author} {\bibfnamefont {M.~H.}\ \bibnamefont {Mu{\~n}oz-Arias}}, \bibinfo {author} {\bibfnamefont {I.~H.}\ \bibnamefont {Deutsch}},\ and\ \bibinfo {author} {\bibfnamefont {P.~M.}\ \bibnamefont {Poggi}},\ }\href@noop {} {\bibfield  {journal} {\bibinfo  {journal} {PRX Quantum}\ }\textbf {\bibinfo {volume} {3}},\ \bibinfo {pages} {010351} (\bibinfo {year} {2022})}\BibitemShut {NoStop}%
\bibitem [{\citenamefont {Hosur}\ \emph {et~al.}(2016)\citenamefont {Hosur}, \citenamefont {Qi}, \citenamefont {Roberts},\ and\ \citenamefont {Yoshida}}]{hosurChaosQuantumChannels2016}%
  \BibitemOpen
  \bibfield  {author} {\bibinfo {author} {\bibfnamefont {P.}~\bibnamefont {Hosur}}, \bibinfo {author} {\bibfnamefont {X.~L.}\ \bibnamefont {Qi}}, \bibinfo {author} {\bibfnamefont {D.~A.}\ \bibnamefont {Roberts}},\ and\ \bibinfo {author} {\bibfnamefont {B.}~\bibnamefont {Yoshida}},\ }\href {https://doi.org/10.1007/JHEP02(2016)004} {\bibfield  {journal} {\bibinfo  {journal} {Journal of High Energy Physics}\ }\textbf {\bibinfo {volume} {2016}},\ \bibinfo {pages} {1} (\bibinfo {year} {2016})}\BibitemShut {NoStop}%
\bibitem [{\citenamefont {Gullans}\ and\ \citenamefont {Huse}(2020)}]{gullans2020dynamical}%
  \BibitemOpen
  \bibfield  {author} {\bibinfo {author} {\bibfnamefont {M.~J.}\ \bibnamefont {Gullans}}\ and\ \bibinfo {author} {\bibfnamefont {D.~A.}\ \bibnamefont {Huse}},\ }\href {https://doi.org/10.1103/PhysRevX.10.041020} {\bibfield  {journal} {\bibinfo  {journal} {Phys. Rev. X}\ }\textbf {\bibinfo {volume} {10}},\ \bibinfo {pages} {041020} (\bibinfo {year} {2020})}\BibitemShut {NoStop}%
\bibitem [{\citenamefont {Zabalo}\ \emph {et~al.}(2020)\citenamefont {Zabalo}, \citenamefont {Gullans}, \citenamefont {Wilson}, \citenamefont {Gopalakrishnan}, \citenamefont {Huse},\ and\ \citenamefont {Pixley}}]{zabalo2020critical}%
  \BibitemOpen
  \bibfield  {author} {\bibinfo {author} {\bibfnamefont {A.}~\bibnamefont {Zabalo}}, \bibinfo {author} {\bibfnamefont {M.~J.}\ \bibnamefont {Gullans}}, \bibinfo {author} {\bibfnamefont {J.~H.}\ \bibnamefont {Wilson}}, \bibinfo {author} {\bibfnamefont {S.}~\bibnamefont {Gopalakrishnan}}, \bibinfo {author} {\bibfnamefont {D.~A.}\ \bibnamefont {Huse}},\ and\ \bibinfo {author} {\bibfnamefont {J.~H.}\ \bibnamefont {Pixley}},\ }\href {https://doi.org/10.1103/PhysRevB.101.060301} {\bibfield  {journal} {\bibinfo  {journal} {Phys. Rev. B}\ }\textbf {\bibinfo {volume} {101}},\ \bibinfo {pages} {060301} (\bibinfo {year} {2020})}\BibitemShut {NoStop}%
\bibitem [{\citenamefont {Seshadri}\ \emph {et~al.}(2018)\citenamefont {Seshadri}, \citenamefont {Madhok},\ and\ \citenamefont {Lakshminarayan}}]{Seshadri_2018}%
  \BibitemOpen
  \bibfield  {author} {\bibinfo {author} {\bibfnamefont {A.}~\bibnamefont {Seshadri}}, \bibinfo {author} {\bibfnamefont {V.}~\bibnamefont {Madhok}},\ and\ \bibinfo {author} {\bibfnamefont {A.}~\bibnamefont {Lakshminarayan}},\ }\href {https://doi.org/10.1103/PhysRevE.98.052205} {\bibfield  {journal} {\bibinfo  {journal} {Physical Review E}\ }\textbf {\bibinfo {volume} {98}},\ \bibinfo {pages} {052205} (\bibinfo {year} {2018})}\BibitemShut {NoStop}%
\bibitem [{\citenamefont {Norcia}\ \emph {et~al.}(2019)\citenamefont {Norcia}, \citenamefont {Young}, \citenamefont {Eckner}, \citenamefont {Oelker}, \citenamefont {Ye},\ and\ \citenamefont {Kaufman}}]{Matthew_opticalclocks}%
  \BibitemOpen
  \bibfield  {author} {\bibinfo {author} {\bibfnamefont {M.~A.}\ \bibnamefont {Norcia}}, \bibinfo {author} {\bibfnamefont {A.~W.}\ \bibnamefont {Young}}, \bibinfo {author} {\bibfnamefont {W.~J.}\ \bibnamefont {Eckner}}, \bibinfo {author} {\bibfnamefont {E.}~\bibnamefont {Oelker}}, \bibinfo {author} {\bibfnamefont {J.}~\bibnamefont {Ye}},\ and\ \bibinfo {author} {\bibfnamefont {A.~M.}\ \bibnamefont {Kaufman}},\ }\href {https://doi.org/10.1126/science.aay0644} {\bibfield  {journal} {\bibinfo  {journal} {Science}\ }\textbf {\bibinfo {volume} {366}},\ \bibinfo {pages} {93} (\bibinfo {year} {2019})}\BibitemShut {NoStop}%
\bibitem [{\citenamefont {Madjarov}\ \emph {et~al.}(2019)\citenamefont {Madjarov}, \citenamefont {Cooper}, \citenamefont {Shaw}, \citenamefont {Covey}, \citenamefont {Schkolnik}, \citenamefont {Yoon}, \citenamefont {Williams},\ and\ \citenamefont {Endres}}]{Madjarov2019_opticalclocks}%
  \BibitemOpen
  \bibfield  {author} {\bibinfo {author} {\bibfnamefont {I.~S.}\ \bibnamefont {Madjarov}}, \bibinfo {author} {\bibfnamefont {A.}~\bibnamefont {Cooper}}, \bibinfo {author} {\bibfnamefont {A.~L.}\ \bibnamefont {Shaw}}, \bibinfo {author} {\bibfnamefont {J.~P.}\ \bibnamefont {Covey}}, \bibinfo {author} {\bibfnamefont {V.}~\bibnamefont {Schkolnik}}, \bibinfo {author} {\bibfnamefont {T.~H.}\ \bibnamefont {Yoon}}, \bibinfo {author} {\bibfnamefont {J.~R.}\ \bibnamefont {Williams}},\ and\ \bibinfo {author} {\bibfnamefont {M.}~\bibnamefont {Endres}},\ }\href {https://doi.org/10.1103/PhysRevX.9.041052} {\bibfield  {journal} {\bibinfo  {journal} {Phys. Rev. X}\ }\textbf {\bibinfo {volume} {9}},\ \bibinfo {pages} {041052} (\bibinfo {year} {2019})}\BibitemShut {NoStop}%
\bibitem [{\citenamefont {Cerezo}\ \emph {et~al.}(2021)\citenamefont {Cerezo}, \citenamefont {Arrasmith}, \citenamefont {Babbush}, \citenamefont {Benjamin}, \citenamefont {Endo}, \citenamefont {Fujii}, \citenamefont {McClean}, \citenamefont {Mitarai}, \citenamefont {Yuan}, \citenamefont {Cincio} \emph {et~al.}}]{cerezo2021}%
  \BibitemOpen
  \bibfield  {author} {\bibinfo {author} {\bibfnamefont {M.}~\bibnamefont {Cerezo}}, \bibinfo {author} {\bibfnamefont {A.}~\bibnamefont {Arrasmith}}, \bibinfo {author} {\bibfnamefont {R.}~\bibnamefont {Babbush}}, \bibinfo {author} {\bibfnamefont {S.~C.}\ \bibnamefont {Benjamin}}, \bibinfo {author} {\bibfnamefont {S.}~\bibnamefont {Endo}}, \bibinfo {author} {\bibfnamefont {K.}~\bibnamefont {Fujii}}, \bibinfo {author} {\bibfnamefont {J.~R.}\ \bibnamefont {McClean}}, \bibinfo {author} {\bibfnamefont {K.}~\bibnamefont {Mitarai}}, \bibinfo {author} {\bibfnamefont {X.}~\bibnamefont {Yuan}}, \bibinfo {author} {\bibfnamefont {L.}~\bibnamefont {Cincio}}, \emph {et~al.},\ }\href@noop {} {\bibfield  {journal} {\bibinfo  {journal} {Nature Reviews Physics}\ }\textbf {\bibinfo {volume} {3}},\ \bibinfo {pages} {625} (\bibinfo {year} {2021})}\BibitemShut {NoStop}%
\bibitem [{dat(2024)}]{data_open_access}%
  \BibitemOpen
  \href {https://doi.org/10.15129/fda72763-74f6-44db-9daf-ff5e28b6ee8e} {\bibinfo {title} {https://doi.org/10.15129/fda72763-74f6-44db-9daf-ff5e28b6ee8e}} (\bibinfo {year} {2024})\BibitemShut {NoStop}%
\bibitem [{\citenamefont {Faccin}\ \emph {et~al.}(2013)\citenamefont {Faccin}, \citenamefont {Johnson}, \citenamefont {Biamonte}, \citenamefont {Kais},\ and\ \citenamefont {Migda{\l}}}]{faccin2013}%
  \BibitemOpen
  \bibfield  {author} {\bibinfo {author} {\bibfnamefont {M.}~\bibnamefont {Faccin}}, \bibinfo {author} {\bibfnamefont {T.}~\bibnamefont {Johnson}}, \bibinfo {author} {\bibfnamefont {J.}~\bibnamefont {Biamonte}}, \bibinfo {author} {\bibfnamefont {S.}~\bibnamefont {Kais}},\ and\ \bibinfo {author} {\bibfnamefont {P.}~\bibnamefont {Migda{\l}}},\ }\href@noop {} {\bibfield  {journal} {\bibinfo  {journal} {Physical Review X}\ }\textbf {\bibinfo {volume} {3}},\ \bibinfo {pages} {041007} (\bibinfo {year} {2013})}\BibitemShut {NoStop}%
\end{thebibliography}%

\pagebreak
\widetext
 
 \newpage 
\begin{center}
\vskip0.5cm
{\Large Supplemental Material}
\end{center}
\vskip0.4cm

\setcounter{section}{0}
\setcounter{equation}{0}
\setcounter{figure}{0}
\setcounter{table}{0}
\setcounter{page}{1}
\renewcommand{\theequation}{S\arabic{equation}}
\renewcommand{\thefigure}{S\arabic{figure}}
\renewcommand{\thesection}{S\arabic{section}}
\renewcommand{\thesection}{S\arabic{section}}
\section{Quantum Fisher Information of Compass States}
\begin{figure}[h] 
        \centering
        \includegraphics[width=1.0\linewidth]{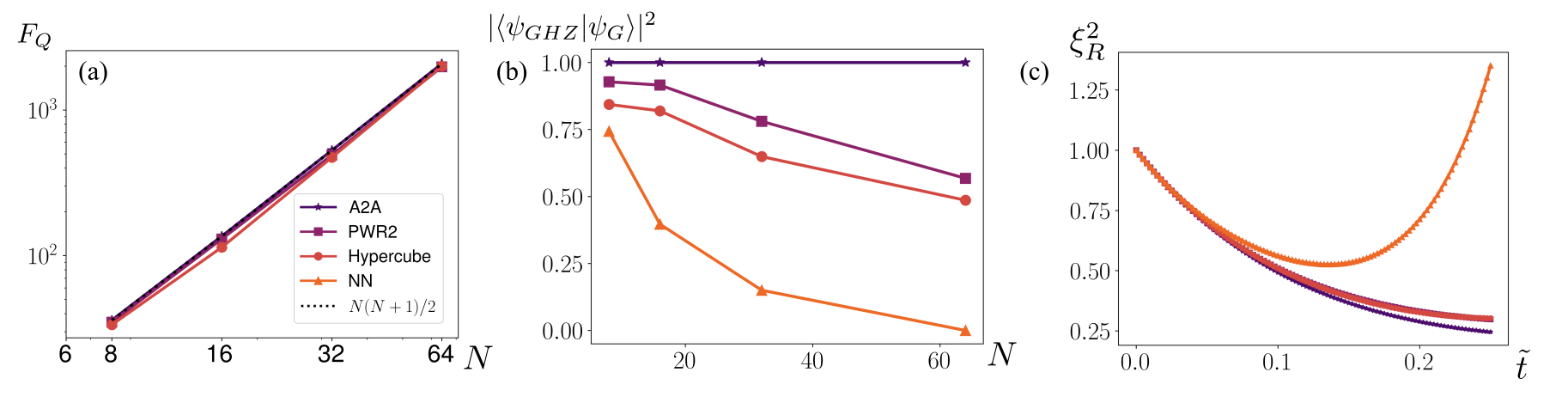}
        \caption{(a) The Quantum Fisher Information of compass states is plotted as a function of different system sizes $N = 8, 16, 32, 64$. The all to all coupling graph creates states that have $F_Q= N(N+1)/2$. Both the PWR2 and hypercube also have similar scaling. The slopes for all-to-all, PWR2, and hypercube are 1.95, 1.93 and 1.97 respectively. The black dashed line indicates the value of $N(N+1)/2$ for different values of $N$.  (b) The overlap of the state with maximum $F_Q$ with the GHZ state plotted as a function of system sizes $N=8, 16, 32, 64$ for different coupling graphs $G$. (c). The Wineland squeezing parameter is plotted as a function of the normalized time $\tilde{t}$ for different coupling graphs for $N=16$. The system is evolved up to a final time $\tilde{t} = 1/\sqrt{N}$. We observe that the A2A, PWR2, and hypercube coupling graphs exhibit significant spin squeezing, with $\xi_R^{2}<1$. In contrast, the NN geometry does not show this behavior.  }
        \label{Appfig:Compass state scaling}
\end{figure}

The states native to the plateau region in the OAT Hamiltonian are called Compass states, characterized by a Quantum Fisher Information $F_Q = N(N+1)/2$. In \figref{Appfig:Compass state scaling}(a), we compare the $F_Q$ of these states for different coupling graphs as a function of the system size. Since the dynamics of the NN geometry do not exhibit a distinct plateau, as shown in \figref{Fig1:Main_Result}(f) of the main text, we exclude it from the following analysis. The sparse graphs (PWR2 and hypercube) demonstrate the same scaling of $F_Q$ with system size as the all-to-all (A2A) coupling graph, indicating Heisenberg scaling.
\section{Overlap with the GHZ state}
We also calculate the overlap of the states with maximum Quantum Fisher Information created by different coupling graphs with the GHZ state. In \figref{Appfig:Compass state scaling}(b), $|\langle \psi_{GHZ}|\psi_{G} \rangle|^2$ is plotted as a function of system size for different coupling graphs $G$. As expected, we see a perfect overlap for the A2A coupling graph. The PWR2 and hypercube coupling graphs have a good overlap with the GHZ state, and in contrast, the NN coupling graph has an overlap that decreases and becomes negligible for larger system sizes. 
\section{Calculation of the Wineland Squeezing Parameter}
To demonstrate that the sparse coupling graphs emulate all-to-all dynamics, we also calculate the Wineland squeezing parameter \cite{Wineland_1994,Wineland_1992} as a function of normalized time $\tilde{t}$, evolving the system up to $\tilde{t} = 1/\sqrt{N}$ for  $N=16$. The Wineland squeezing parameter is defined as:
\begin{align}
\label{Eq:Wine_Spin_squeezing}
\xi _{R}^{2} = \frac{N (\Delta \hat{J}_{\perp})^2}{|\langle \hat{J}_{\boldsymbol{s}} \rangle |^2} ~,
\end{align}
where $\boldsymbol{s}$ is the mean-spin direction and $\perp$ denotes a perpendicular direction to the mean-spin direction. If $\xi _{R}^{2} < 1$, the state is said to be spin squeezed along the $\perp$ axis. Both the PWR2 and hypercube coupling graphs exhibit similar behavior in terms of the squeezing parameter $\xi _{R}^{2}$, whereas the nearest neighbour coupling graph shows significantly less squeezing as shown in \figref{Appfig:Compass state scaling}(c).
\section{Stroboscopic evolution to create states with Heisenberg scaling}
\begin{figure}
    \centering
    \includegraphics[width=0.7\linewidth]{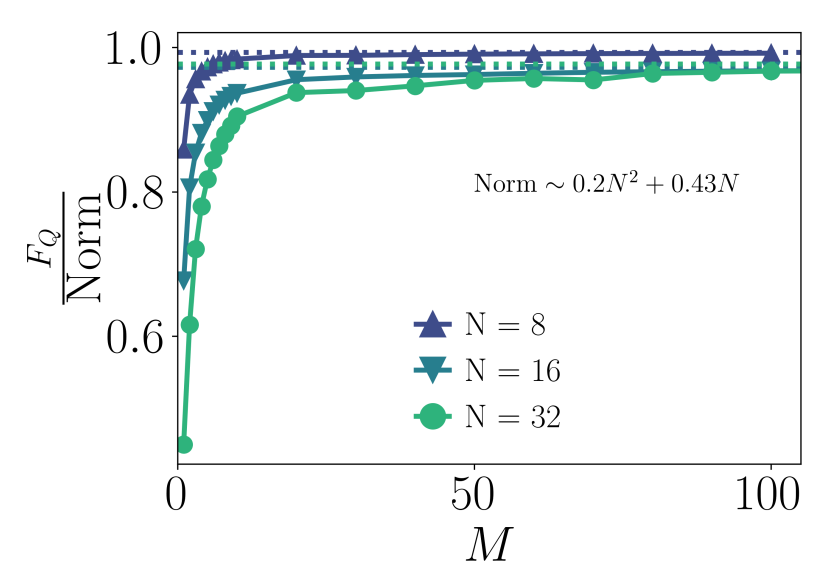}
    \caption{The normalised QFI, $\frac{F_Q}{\textrm{Norm}}$, is shown as a function of the number of iterations $M$, where the initial $x$-polarised state is evolved up to $t = 1/\sqrt{N}$ under the $XY$ Hamiltonian, with $N$ representing the system size. The dashed lines of the corresponding colors represent the values of the normalised QFI obtained from continuous time evolution.}
    \label{fig:strob_sqrt}
\end{figure}

In the main text, we demonstrate how the maximum value of the QFI varies with the number of iterations, $M$ (see \figref{Fig4:Experimental} (b)), and observe a sub-linear scaling of $M$ with system size. However, it is possible to generate states that exhibit Heisenberg scaling in significantly shorter times than the characteristic time $t^{*}$ \cite{PezzeSmerzi2014}. Here, we show that by evolving the initial $x$-polarised state under the $XY$ Hamiltonian to a time $t=1/\sqrt{N}$, we can produce states with Heisenberg scaling. This is illustrated in \figref{fig:strob_sqrt}, where the normalised QFI is plotted as a function of the number of iterations. The QFI is normalised by a factor of $N^2$, highlighting the presence of Heisenberg scaling.  
\section{Convergence analysis for the MPS simulations}
\begin{figure}[h] 
        \centering
        \includegraphics[width=0.90\linewidth]{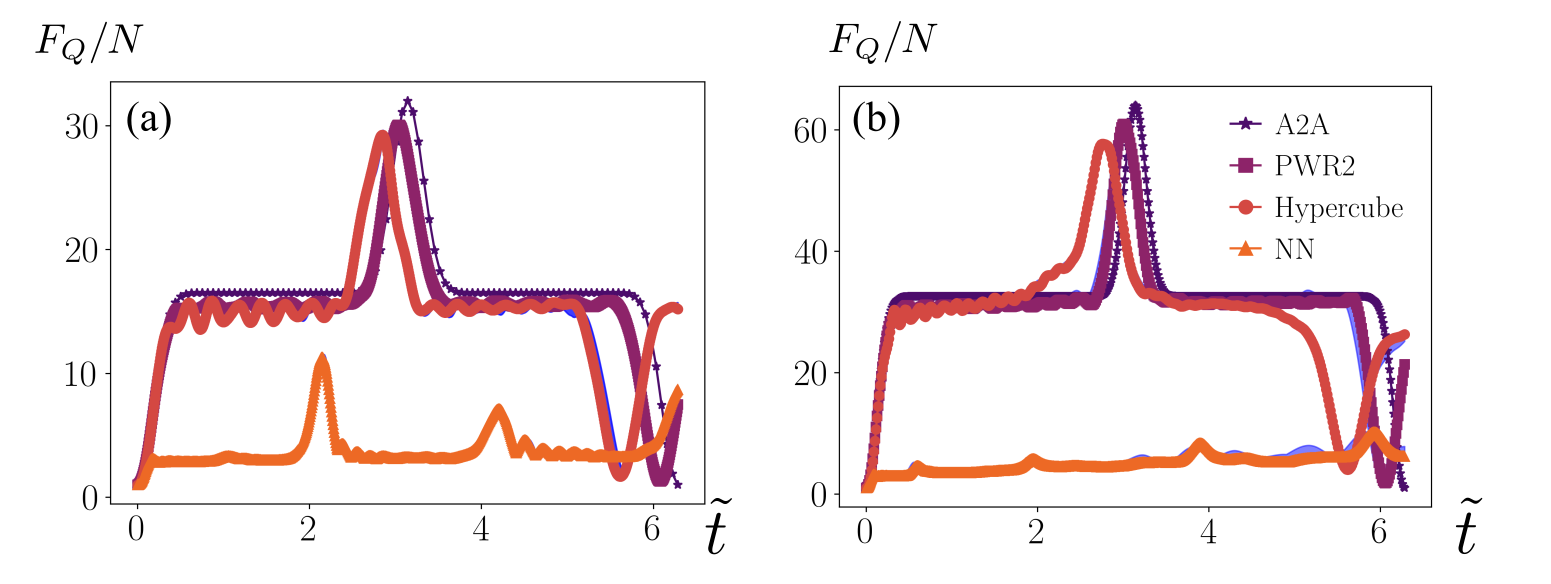}
        \caption{Quantum Fisher Information $F_Q$ is plotted as a function of the normalized time $\tilde{t}$ for all the different coupling graphs for (a) $N=32$ and $N=64$. For conducting convergence analysis, for (a) $N=32$, the results in the main text are for the following parameters $[D=256, \chi_{0}dt =10^{-2}, \epsilon = 10^{-13}]$. We have added maximal and minimal deviation as a shaded region around this result for $[D=256, \chi_{0}dt =10^{-1}, \epsilon = 10^{-13}]$, $[D=512, \chi_{0}dt =10^{-2}, \epsilon = 10^{-13}]$, $[D=256, \chi_{0}dt =10^{-2}, \epsilon = 10^{-9}]$. On the other hand, to ensure convergence for (b) $N=64$, the result shown in the main text are for the following parameters $[D=512, \chi_{0}dt =10^{-2}, \epsilon = 10^{-13}]$. We have added the maximal and minimal deviation as a shaded region around this result for $[D=512, \chi_{0}dt =10^{-3}, \epsilon = 10^{-13}]$, $[D=512, \chi_{0}dt =10^{-2}, \epsilon = 10^{-9}]$, and $[D=256, \chi_{0}dt =10^{-2}, \epsilon = 10^{-13}]$ for all the coupling graphs except hypercube. For hypercube, we use $D=400$, instead of $D=256$. If the shading is not discernible, the numerical error is below the linewidth shown. }
        \label{Appfig:Convergence}
\end{figure}
To ensure the convergence of our results, we vary all the free parameters in our MPS simulations. As mentioned in the main text, the results for $N=32$ are for the following parameters $[D=256, \chi_{0}dt =10^{-2}, \epsilon = 10^{-13}]$ and for $N=64$ are $[D=512, \chi_{0}dt =10^{-2}, \epsilon = 10^{-13}]$. For $N=32$, we have added maximal and minimal deviation as a shaded region around this result for $[D=256, \chi_{0}dt =10^{-1}, \epsilon = 10^{-13}]$, $[D=512, \chi_{0}dt =10^{-2}, \epsilon = 10^{-13}]$, $[D=256, \chi_{0}dt =10^{-2}, \epsilon = 10^{-9}]$ as shown in \figref{Appfig:Convergence}. 
For $N=64$, we have added maximal and minimal deviation as a shaded region around this result for $[D=512, \chi_{0}dt =10^{-3}, \epsilon = 10^{-13}]$, $[D=512, \chi_{0}dt =10^{-2}, \epsilon = 10^{-9}]$, and $[D=256, \chi_{0}dt =10^{-2}, \epsilon = 10^{-13}]$ for all the coupling geometries except the hypercube, where instead we use $[D=512, \chi_{0}dt =10^{-3}, \epsilon = 10^{-13}]$, $[D=512, \chi_{0}dt =10^{-2}, \epsilon = 10^{-9}]$, and $[D=400, \chi_{0}dt =10^{-2}, \epsilon = 10^{-13}]$ for the convergence analysis. If the shading is invisible, the numerical error is below the line width. 

\section{Excitation spectra for the isotropic Heisenberg model in sparse coupling graphs}

Consider the generalized one-axis-twisting Hamiltonian,

\begin{equation}
H_{{\rm gOAT}} = \chi_0 J_z^2 - \frac{1}{4}\sum\limits_{ij} \chi_{ij} \vec{\sigma_i}.\vec{\sigma_j} ~.
\end{equation}

It is straightforward to check that $[H_{{\rm gOAT}},\mathbf{J}^2]=[H_{{\rm gOAT}},J_z]=0$, and thus $H_{{\rm gOAT}}$ shares a basis of eigenvectors with $\mathbf{J^2}$ and $J_z$. Moreover, both terms of $H_{{\rm gOAT}}$ commute, and so to analyze the energy gap between subspaces of defined $\mathbf{J}$ we focus our analysis on the isotropic Heisenberg Hamiltonian,
\begin{equation}
    H_H = - \frac{1}{4}\sum\limits_{ij} \chi_{ij} \vec{\sigma_i}.\vec{\sigma_j} = -\sum\limits_{ij} \chi_{ij} h_{(i,j)} ~,
\end{equation}

\noindent which has a continuous SU$(2)$ symmetry, and where we have defined
\begin{equation}
    h_{(i,j)}=\frac{1}{4}\vec{\sigma_i}.\vec{\sigma_j} = \frac{1}{2}\left(\sigma_i^+ \sigma_j^- + \sigma_i^- \sigma_j^+\right) + \frac{1}{4}\sigma_i^z \sigma_j^z ~. 
\end{equation}

The ground states of this model correspond to all spins pointing along the same direction, for instance 
\begin{equation}
    \Ket{\Phi_0} = \Ket{0}^{\otimes N} ~,
\end{equation}
\noindent for which the associated energy is
\begin{equation}
    E_0 = -\frac{1}{4}\sum\limits_{ij} \chi_{ij} ~.
\end{equation}

We can construct excitations from this base state by flipping a single spin, which leads to the family of states
\begin{equation}
    \Ket{\phi_k} = \Ket{00\ldots 0 1_k 0 \ldots },\ k=1,\ldots,N ~.
\end{equation}

The Heisenberg interaction acts on this state as
\begin{equation}
    h_{(i,j)}\Ket{\phi_k} = \frac{1}{2}\left(\delta_{jk}\Ket{\phi_i} + \delta_{ik}\Ket{\phi_j}\right)+\frac{1} {4}\left(1-2(\delta_{ik}+\delta_{jk})\right)\Ket{\phi_k} ~.
\end{equation}

Using that $\chi_{ij}=\chi_{ji}$ we can prove that
\begin{equation}
    H_H \Ket{\phi_k} = \left(E_0 + \sum\limits_{j} \chi_{kj}\right)\Ket{\phi_k} - \sum\limits_i \chi_{ik}\Ket{\phi_i} ~.
    \label{eq:hami_phi_k}
\end{equation}

\subsection{One-dimensional graphs}

The eigenstates of $H_H$ can be obtained exactly when the coupling graph shows translational invariance. In 1D with periodic boundary conditions, this means that the coupling strengths can expressed as a function of
\begin{equation}
    \chi_{ij}\equiv \chi(|i-j|) ~,
\end{equation}
\noindent which obeys $\chi(x)=\chi(x\pm N)$, $\chi(0)=0$. For this case the eigenstates are spin waves \cite{Pires_spinwavetheory,stancel2009spin,kittel1963quantum}, i.e.

\begin{equation}
    \Ket{\Phi_q} = \frac{1}{\sqrt{N}}\sum\limits_{k=1}^N e^{ik\frac{2\pi}{N}q}\Ket{\phi_k},\ \text{with}\ q=1,\ldots,N-1 ~,
\end{equation}

\noindent for which we have

\begin{equation}
    H_H\Ket{\Phi_q} = E_0\Ket{\Phi_q}+\frac{1}{\sqrt{N}}\sum\limits_k e^{ik\frac{2\pi}{N}q}\left(\sum\limits_j \chi(k-j)\left(1-e^{i(j-k)\frac{2\pi}{N}q}\right)\right)\Ket{\phi_k} ~.
    \label{eq:gap_H_on_Phi}
\end{equation}

Using translational invariance and periodic boundary conditions, we can derive an expression for the gaps $\Delta_1(q)=E_q-E_0$, which read
\begin{align}
    N\ \text{even}: & \ \Delta_1(q) = 2\sum\limits_{m=1}^{\frac{N}{2}} \chi(m) \left(1-\cos\left(\frac{2\pi}{N}mq\right)\right) - \chi\left(\frac{N}{2}\right)(1-e^{i\pi q}) ~, \label{eq:gap_even} \\
    N\ \text{odd}: & \ \Delta_1(q) = 2\sum\limits_{m=1}^{\frac{N-1}{2}} \chi(m) \left(1-\cos\left(\frac{2\pi}{N}mq\right)\right) ~.
\end{align}

In many cases, the lowest-energy excitation will correspond to $q=1$, but this is not universally the case as it depends on the coupling graph.\\

\paragraph{PWR2 graph}.

For the PWR2 graph with periodic boundary conditions, we have that
\begin{equation}
    \chi(m) = \chi_0, \ \text{if}\ m=2^r\ \text{or}\ N-m = 2^r,\ r= 0, 1,\ldots ~.
\end{equation}

Restricting to even $N$, we find that the smallest gap happens for $q=N/2$. In this case,
\begin{equation}
  \Delta_{\text{PWR2}} = \sum\limits_{r=0}^{\log_{2}(N)-1} 2\chi_0 \left(1-\cos\left(\pi 2^r\right)\right) - 2\chi\left(\frac{N}{2}\right)(1-e^{i\pi\frac{N}{2}}) = 4\chi_0 ~,
\end{equation}

\noindent where the result follows from the fact that only the $r=0$ term survives in the sum, and the second term is 0 since either $\chi(N/2)=0$ for $N/2$ odd or even but not a power of two, or $e^{i\pi N/2}=1$ for $N/2$ equal a power of two larger than 1. The result then indicates that the energy gap separating the collective states from the rest of the spectrum is finite for all $N$, and actually independent of system size.\\

\paragraph{Power-law decaying interactions.} The case of systems with all-to-all interactions which decay as a power of the distance, i.e.
\begin{equation}
    \chi_{ij} = \frac{\chi_0}{|i-j|^\alpha}, \alpha \geq 0 ~,
\end{equation}

\noindent was studied by Perlin \textit{et al.} in Ref. \cite{perlin2020}. For this case the smallest gap is obtained for $q=1$. Ref.  \cite{perlin2020} showed that, for $N\gg 1$, the gap scales as
\begin{equation}
    \Delta_{\text{AD}} \sim N^{1-\alpha} ~. 
\end{equation}
As a result, interactions with $\alpha \geq 1$ will show gaps that decay  with $N$, but interactions with $\alpha\leq 1$ will have gaps that increase (or be asymptotically constant) with $N$ and thus robustly preserve collective states. \\

\paragraph{All-to-all couplings.} The all-to-all case is contained in the power-law description when $\alpha=0$. The gap can be worked out exactly from Eq. (\ref{eq:gap_even}) and gives

\begin{equation}
    \Delta_{\text{A2A}} = 2\chi_0\sum\limits_{m=1}^{\frac{N}{2}} \left(1-\cos\left(\frac{2\pi}{N}mq\right)\right) - 2\chi_0 = \chi_0 N ~,
\end{equation}

\noindent which shows the expected $\sim N$ scaling.\\

\paragraph{Nearest-neighbour coupling} For nearest neighbour coupling with periodic boundary conditions,
\begin{equation}
    \chi(m) = \chi_0(\delta_{m,1} + \delta_{m,N-1}) ~.
\end{equation}

Assuming even $N$, we again get that the smallest gap is obtained for $q=1$. The gap reads 
\begin{equation}
    \Delta_{\text{NN}} = 2\chi_0\left(1-\cos\left(\frac{2\pi}{N}\right)\right)\sim 4\pi^2 N^{-2} ~, 
\end{equation}
\noindent which decays as $N$ increases.

\subsection{Hypercube graph}

The calculation of the hypercube graph in this context requires some care because this graph is not translationally-invariant when seen as a 1D connectivity graph. Naturally, however, the hypercube is "translationally invariant" by definition in $D=\log_{2}(N)$ dimensions, since each direction supports only 2 particles. Thus, we expect the excited states to have the form of $D$-dimensional spin-wave states. To construct these explicitly, for a given site $k=1,2,\ldots,N$ we can construct a $D$-dimensional vector from the binary digits of $k$, i.e.
\begin{equation}
    \vec{B}_k = (b_0,b_1,\ldots,b_{D-1}) ~.
\end{equation}

From these vector we can express the Hamming distance between two sites as $d_H(i,j)=|\vec{B}_i - \vec{B}_j|_1$. The Hypercube graph is such that $\chi_{ij}=\chi_0 \Leftrightarrow d_H(i,j)=1$, and $\chi_{ij}=0$ otherwise. In this picture, we construct the spin-wave state as
\begin{equation}
    \Ket{\Phi_{\vec{q}}} = \sum\limits_{k=1}^N e^{i \frac{2\pi}{N}\vec{B}_k.\vec{q}} \Ket{\phi_k} ~,
\end{equation}
\noindent where $\vec{q}$ is a now a $D$-dimensional spin-wave vector. Since each direction supports only two particles, each wavenumber can only take up the values $q_i=0,1$. The action of $H_H$ on this state is analogous to Eq. (\ref{eq:gap_H_on_Phi})
\begin{equation}
    H_H \Ket{\Phi_{\vec{q}}}=E_0 \Ket{\Phi_{\vec{q}}} + \frac{1}{\sqrt{N}}\sum\limits_k e^{i \frac{2\pi}{N}\vec{B}_k.\vec{q}} \left[\sum\limits_j \chi\left(|\vec{B}_j - \vec{B}_k|_1\right)\left(1- e^{i \frac{2\pi}{N}(\vec{B}_j-\vec{B}_k).\vec{q}}\right)\right]\Ket{\Phi_{\vec{q}}} ~.
\end{equation}
As established before, for a given site there are $D$ other sites connected to it, and thus $D$ choices of $j$ such that $|\vec{B}_j - \vec{B}_k|_1$. The expression in brackets, which corresponds to the gap $\Delta_{\text{HYP}}$ can then be written as

\begin{equation}
\Delta_{\text{HYP}}=\chi_0\sum\limits_{l=1}^D\left(1-e^{i\pi q_l}\right),\ \text{where}\ q_l = 0, 1 ~.
\end{equation}

Setting all $q_l=0$ recovers the ground state. So, the first excitation corresponds to all $q_l=0$ except one $q_r=1$, leading to a gap
\begin{equation}
    \Delta_{\text{HYP}}=\chi_0(D-(D-2))=2\chi_0 ~,
\end{equation}

\noindent which is independent of $N$ as the case of the PWR2 coupling graph.

\subsection{General graphs}
We note that Eq. (\ref{eq:hami_phi_k}) implies that $H_H$ in the single excitation subspace can be written as
\begin{equation}
    H_H = E_0 \mathbb{I} + D - A ~,
\end{equation}
where $D\rightarrow D_{ij} = \sum\limits_i \chi_{ij}$ is the degree matrix of the coupling graph and $A\rightarrow A_{ij}=\chi_{ij}$ is its adjacency matrix. The matrix $L=D-A$ is the Laplacian of the graph, which is the generator of motion for a (classical) stochastic random walk on such graphs \cite{faccin2013} Thus, the smallest gap of $L$ dictates the timescale required for the random walk to reach its equilibrium distribution. This connection between the energy gap protecting collective dynamics and the random walk gap can be used to systematically explore the use of any graph for the generation of metrologically-useful states. We leave this task for future work.

\section{Experimental implementation using neutral atoms in tweezer arrays}
In the main text we presented a stroboscopic protocol to implement the XY spin interactions according to a hypercube graph using neutral atoms in reconfigurable tweezer arrays. In this section we provide further details about this protocol, and expand our proposal to allow implementation of the PWR2 model as well.

\subsection{Hypercube graph}

As shown in Fig. 4 of the main text, the proposed protocol has three main ingredients: 
\begin{enumerate}
    \item dynamical optical tweezer reconfiguration which allows to shuffle the position of the atoms,  
    \item local rotations $R_\alpha$ around the $\alpha=x,y$ axis by a fixed angle $\pi/2$,
    \item implementation of $zz$ or Ising interaction according to $H_{zz}$ for a time $dt/2=t^\ast/(2M)\equiv 4\theta$. This is equivalent to the parallel implementation of $N/2$ Ising gates between atoms $2\nu$ and $2\nu-1$ ($\nu=1,\ldots, N$)
    \begin{equation}
        U_{zz} = e^{-i \sigma_1^z \sigma_2^z \theta} ~.
    \label{eq:SM_ising_gate}
    \end{equation}
\end{enumerate}

We consider the atoms as three-level systems $\{\Ket{0},\Ket{1},\Ket{r}\}$ where the qubit is encoded in $\{\Ket{0},\Ket{1}\}$, and $\Ket{r}$ is a Rydberg state. Local rotations are readily achieved by using microwave or Raman pulses. To entangle the atoms, we leverage the Rydberg blockade mechanism to implement controlled phase gates $U_{\rm{CZ}}(\phi)$, following known proposals like \cite{levine2019parallel}  and \cite{fromonteil2024}. We note that Ising interactions are naturally achievable if one considers a pseudo-spin-1/2 encoded in the $\Ket{1}$ and $\Ket{r}$ state (see \cite{Bornet_2023,Labuhn2016} for example). In our case, however, we require the atoms to be in the hyperfine manifold while the atoms are being repositioned, as the Rydberg state is not trappable.

In the computational basis, an arbitrary CPHASE gate has the form
\begin{equation}
    U_{\rm{CZ}}(\phi) = \mathrm{diag}[1,1,1,e^{-i\phi}] ~.
\end{equation}
We can compile the required Ising gate, cf. Eq. (\ref{eq:SM_ising_gate}) from the CPHASE and global rotations $R_z(\varphi)=e^{-i(\sigma_1^z + \sigma_2^z)\varphi/2}$ by noting that

\begin{equation}
    e^{i\alpha}R_z(\varphi)U_{\rm{CZ}}(\phi) = \mathrm{diag}[e^{-i(\varphi-\alpha)},e^{i\alpha},e^{i\alpha},e^{i(\alpha+\varphi-\phi)}] ~.
\end{equation}
Equating this to $U_{zz}(\theta)$ leads to the solution 
\begin{equation}
    \phi = 4\theta\ \mathrm{and}\ \varphi = 2\theta ~,
\end{equation}
where we reiterate that $\theta = dt/8$.

With this procedure in place, we can now visualize each round of the stroboscopic protocol following Fig. \ref{fig:SM_hyp_shuffle}. Note that the atoms are positioned such that only two of them sit within a blockade radius, in such a way that global Rydberg driving leads to the parallel implementation of CPHASE gates. After the sequence of one and two-qubit gates depicted in the figure, the tweezer-assisted Faro shuffle is performed, and the sequence is repeated. In a single Trotter step, this procedure is performed a total of $\log_{2}(N)$ times. This leads to a total of $N\log_{2}(N)$ CPHASE gates per Trotter step. 
Given state-of-the-art techniques, the free shuttling of atoms in AOD tweezers has negligible fidelity loss and therefore does not impact our gate operations \cite{Bluvstein_2022,Bluvstein_2023}. While the currently demonstrated fidelity of $CZ$ gate is $99.5\%$, with realistic estimates up to $99.8\%$ \cite{Evered2023,Pelegri_2022}, some errors can be mitigated through alternative experimental techniques, making $99.9\%$ a realistic target \cite{Bluvstein_2023,Evered2023,Jandura_2022,Pagano_2022}. Single-qubit gates executed globally have a fidelity of $99.99\%$ \cite{Bluvstein_2023,nikolov2023}. Our primary focus is the generation of spin-squeezed states that exhibit Heisenberg scaling in the quantum Fisher information. For a system of $N=16$, as shown in Fig. S2 in our supplementary material, reaching such a state requires approximately $M \approx 10$ Trotter steps using the hypercube coupling graph. Each Trotter step has $N\log_{2}N$ $CZ$ gates, resulting in a total of $M \times N\log_{2}N$ two-qubit gates. Additionally, each two-qubit gate is accompanied by two global single-qubit gates, leading to a total of $M \times 2N\log_{2}N$ single-qubit operations. Given the errors mentioned above, for $N=16$, we calculate the overall many-body state fidelity to be $46.37\%$. Using a slightly lower theoretical estimate of $99.8\%$ for the 
CZ gate fidelity, consistent with recent literature\cite{Evered2023,Pelegri_2022}, this fidelity reduces to $24.4\%$. While actually predicting the achievable values of spin squeezing metrics is challenging numerically for a system of this size in the presence of decoherence and noise, this estimate yielding a high many-body fidelity is a good indicator of the feasibility of our approach. In this simplified estimate, we do not account for error correlations, as the gates are applied in parallel. However, a detailed analysis of these correlations is beyond the scope of this work.

\begin{figure}
    \centering
    \includegraphics[width=0.7\linewidth]{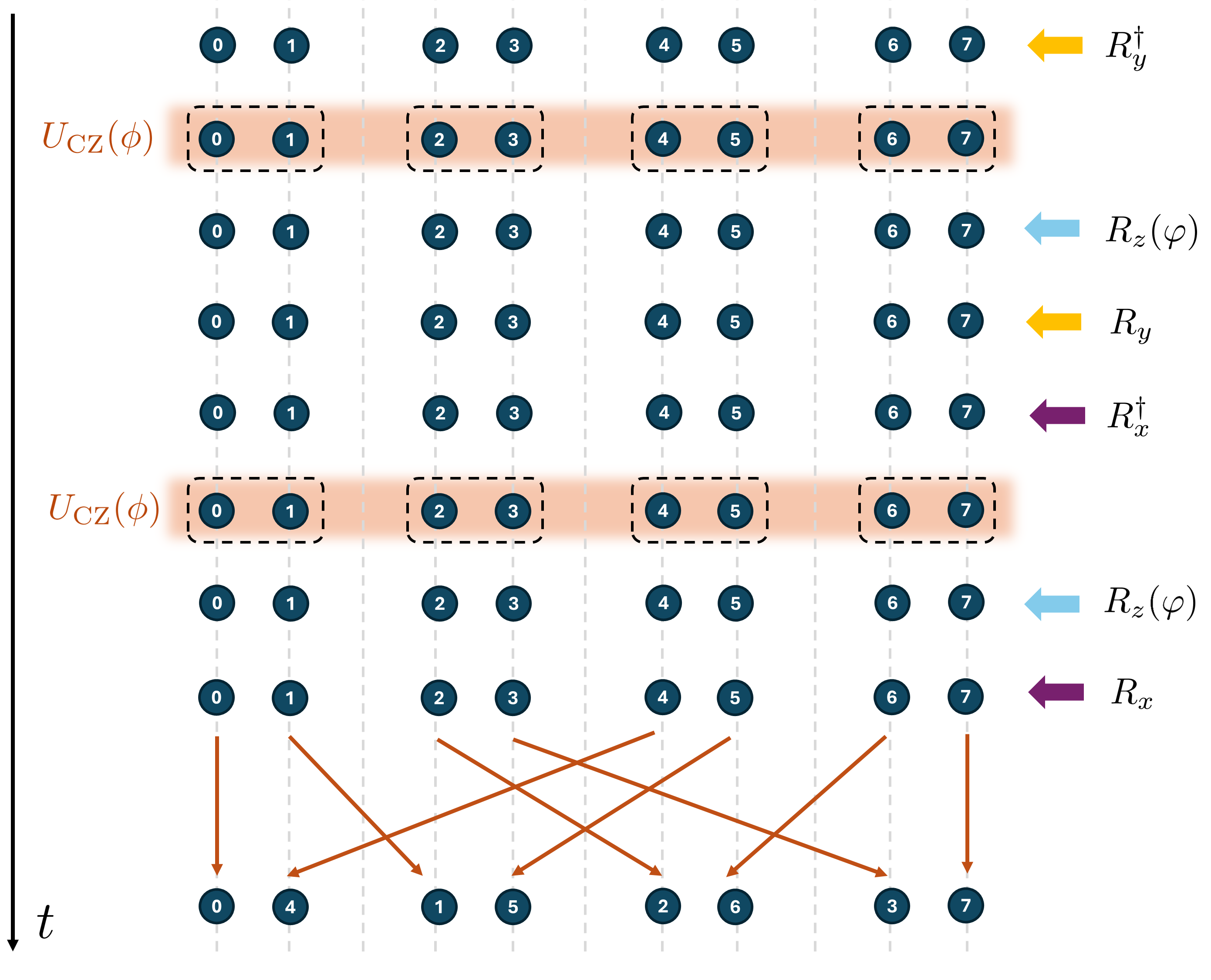}
    \caption{Schematic of a single layer of the stroboscopic protocol depicted in Fig. 4 of the main text. Vertical dashed lines indicate tweezer positions, and each row is a time step of the protocol. The $H_{zz}$ interaction is obtained by combining CPHASE gates $U_{\rm CZ}(\phi)$ and local z-rotations $R_z(\varphi)$ where angles are chosen depending on the length of the Trotter step $dt$. Note that sequences of multiple local rotations (for instance $R_z\rightarrow R_y\rightarrow R_x^\dagger$) could in principle be compiled into a single rotation around a tilted axis.}
    \label{fig:SM_hyp_shuffle}
\end{figure}
\subsection{Powers of Two (PWR2) graph}
In this section, we briefly outline how to dynamically generate a PWR2 sparse coupling graph in a manner very similar to that outlined in the main text for the hypercube. The general setup is shown schematically in \figref{Fig4:Experimental}(a) in the main text. As discussed in the last section, the stroboscopic nearest-neighbour interactions according to \eqeqref{Eq:Strob_ZZ_Hamiltonian} may be implemented using global rotations and arbitrary controlled phase gates.
\begin{figure}[ht!] 
        \centering      
        \includegraphics[width=0.7\linewidth]{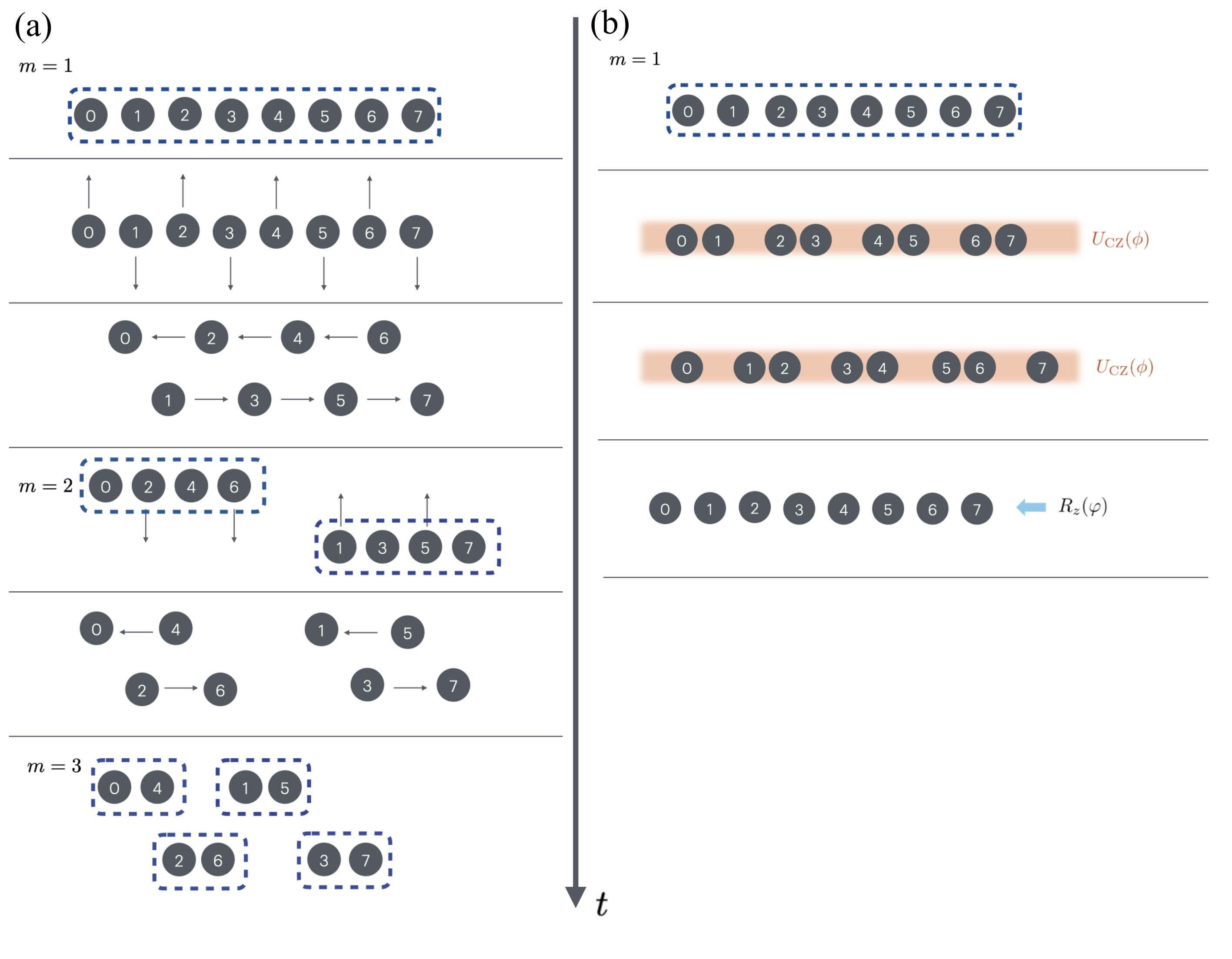}
        \caption{Schematic representation of a rearrangement protocol to dynamically implement a PWR2 coupling graph using neutral atoms (grey circles) in tweezer arrays, shown here for $N=8$ atoms. (a) The protocol involves selectively moving atoms vertically (indicated by arrows) followed by 1D row compressions to compactly rearrange the qubits. The dotted boxes highlight regions where the $H_{zz}$ interactions are applied. (b) During each stage $m \in \{1,\ldots, \log_{2}(N)\}$, the $H_{zz}$ interactions are realised by using two layers of arbitrary CPHASE gates, with 1D row moves used to isolate pair-wise interactions, followed by single qubit rotations.
        }
        \label{Appfig:PWR2shuffling}
\end{figure}
We propose an iterative re-arrangement procedure for the PWR2 model by selectively moving atoms vertically followed by 1D row compressions to compactly rearrange the atomic qubits as illustrated in \figref{Appfig:PWR2shuffling}(a) for a $N=8$ atom system requiring $\log_{2}(N)=3$ shuffling stages. The nearest-neighbour Ising interactions at each stage $m \in \{1, \ldots \log_{2}(N)\}$, indicated by the dashed boxes around groups of qubits, are applied according to the protocol shown in \figref{Appfig:PWR2shuffling}(b). 

For the PWR2 model Ising interactions are realised by applying two layers of CPHASE gates, with 1D row moves used to isolate pair-wise interactions, followed by a single qubit rotation pulse. The two qubits at the end of the chain only have a single neighbour each, requiring half the rotation angle which can be implemented either using local addressing of the end qubits or sequential global pulses to inner and outer qubits. For the final step with $m=\log_{2}(N)$ only a single layer of CPHASE gates is required as the Ising interaction is only applied to atom pairs in this stage.

After this final step, the atoms are returned back to the original layout of stage $m=1$, which concludes one time-step in the evolution under a PWR2 coupling geometry. For periodic boundary conditions the atoms would have to be arranged in a circular geometry. 
\end{document}